\documentclass[reprint,superscriptaddress,bibnotes,amsmath,amssymb,aps,pra,floatfix]{revtex4-2}

\usepackage{graphicx}
\usepackage{dcolumn}
\usepackage{bm}
\usepackage{upgreek}
\usepackage{physics}
\usepackage{hyperref}
\usepackage{xcolor}
\usepackage{sidecap}
\usepackage[T1]{fontenc}
\DeclareUnicodeCharacter{2215}{ }
\DeclareGraphicsExtensions{.pdf,.png}

\begin{document}

\title{Optical Spin Initialisation and Readout with a Cavity-Coupled Quantum Dot in an In-Plane Magnetic Field}

\author{Samuel J. Sheldon}
\affiliation{School of Mathematical and Physical Sciences, University of Sheffield, Sheffield S3 7RH, United Kingdom}
\author{Alistair J. Brash}
\affiliation{School of Mathematical and Physical Sciences, University of Sheffield, Sheffield S3 7RH, United Kingdom}
\author{Maurice S. Skolnick}
\affiliation{School of Mathematical and Physical Sciences, University of Sheffield, Sheffield S3 7RH, United Kingdom}
\author{A. Mark Fox}
\affiliation{School of Mathematical and Physical Sciences, University of Sheffield, Sheffield S3 7RH, United Kingdom}
\author{Jake Iles-Smith}
\affiliation{School of Mathematical and Physical Sciences, University of Sheffield, Sheffield S3 7RH, United Kingdom}
\affiliation{Department of Physics and Astronomy, The University of Manchester, Oxford Road, Manchester M13 9PL, United Kingdom}

\date{\today}

\begin{abstract}
    The spin of a charged semiconductor quantum dot (QD) coupled to an optical cavity is a promising candidate for high fidelity spin-photon
    interfaces; the cavity selectively modifies the decay rates of optical transitions such that spin initialisation, manipulation, and readout are all possible in a single magnetic field geometry.
    By performing cavity QED calculations, we show that a cavity with a single, linearly-polarised mode can simultaneously support both high-fidelity optical spin initialisation and readout in a single, in-plane (Voigt geometry) magnetic field.
    Furthermore, we demonstrate that single mode cavities always outperform bi-modal cavities in experimentally favourable driving regimes.
    Our analysis, when combined with established methods of control in a Voigt geometry field, provides optimal parameter regimes for high-fidelity initialisation and readout, and coherent control in both cavity configurations, providing insights for the design and development of QD spin-photon interfaces as the basis of quantum network nodes and for the generation of photonic graph states.
\end{abstract}

\maketitle

\section{Introduction}

\begin{figure*}[ht!]
    \centering
    \includegraphics[width = 17.5cm]{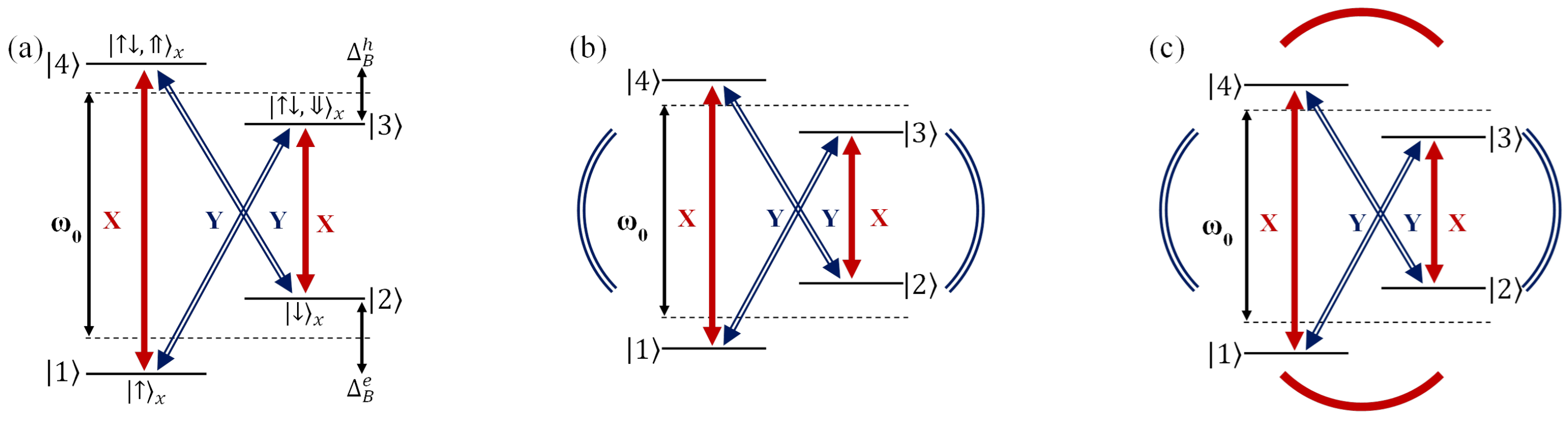}
    \caption{(a) An energy level diagram of a negatively charged quantum dot (QD) in an applied Voigt geometry magnetic field. The degeneracy of the ground and excited eigenstates, written in the basis along the magnetic field axis parallel to the $x$ axis, is lifted by the Zeeman interaction. All four transitions are equally allowed leading to the formation of two $\Lambda$-systems ($1\rightarrow4\rightarrow2$ and $2\rightarrow3\rightarrow1$). The vertical transitions possess the orthogonal linear polarisation to the diagonal transitions.  (b) A $Y$-polarised single-mode cavity coupled to the $Y-$polarised diagonal transitions of a charged-QD in a Voigt geometry magnetic field. (c) A bi-modal cavity with two orthogonal, linearly polarised cavity modes coupled to a charged-QD in a Voigt geometry magnetic field. In b,c the curved lines indicate the configuration of the cavity mode(s), and the line style (solid or compound) and colour indicates the transitions to which the cavity mode couples.}
    \label{fig:cavity_config}
\end{figure*}

Experiments have demonstrated that the spin states of single electrons confined to quantum dots (QDs) are long-lived~\cite{kroutvar2004,gillard2021fundamental}, with coherence times exceeding $\upmu$s~\cite{Press2010,Bechtold2015}, making them promising candidates for an efficient light-matter interface~\cite{kimble2008quantum}.
Furthermore, quantum logic gates may be implemented on ps timescales using ultra-fast optical manipulation of spin-states~\cite{Press2010}.
Combined, these disparate timescales allow many gate operations to be performed within a single lifetime of the charge state.
However, for spin-photon interfaces to be used in optical quantum technologies, such as efficient generation of entangled photonic graph states~\cite{PhysRevLett.103.113602,denning2017protocol,scerri2018cluster,Schwartz434cluster,cogan2023deterministic,Coste:2022hfm,PhysRevLett.128.233602,meng2023photonic} and  spin-photon logic~\cite{Sun2016,Sun57} required for many long-range secure quantum network protocols~\cite{PhysRevX.10.021071,Munro2012}, it is necessary to prepare, control, and readout the single spin states with a high fidelity.

Ordinarily, combining optical spin control and read-out requires the use of orthogonal magnetic field geometries: spin control relies on 
the in-plane Voigt geometry (Fig.~\ref{fig:cavity_config}a) to provide an effective coupling between the spin ground states~\cite{atature2006unity}, while read-out drives a so-called \emph{cyclic transition}~\cite{Kim2008}\footnote{A cycling transition is one which returns the system to its original spin state in a cyclical fashion.} in an out-of-plane Faraday field to produce a detectable signal, ideally within the `single-shot' limit~\cite{PhysRevLett.112.116802,antoniadis2023cavity}.
However, optical spin read-out can also be achieved in the Voigt geometry through polarisation-dependent Purcell enhancement of the optical transitions introduced by coupling to nanophotonic structures~\cite{PhysRevA.94.012307,PhysRevApplied.9.054013,PhysRevLett.126.013602}.

In this work we investigate the impact of coupling between a single charged QD in a Voigt geometry magnetic field, and a single cavity supporting either one or two confined modes (here on referred to as a single-mode or bi-modal cavity respectively), on the spin initialisation and readout fidelity.
We show that, depending on the linewidths and detunings of the cavity modes, a bi-modal cavity may enhance all optical transitions present in a Voigt geometry field, reducing the
effectiveness of the quasi-cycling transition. 
While careful selection of the bi-modal cavity parameters may mitigate these effects, the additional enhancement of the QD transitions result in different cavity parameter requirements for initialisation and readout.
Thus, when coupled to a bi-modal cavity we show there is no single set of cavity parameters that simultaneously results in both a good optical initialisation and readout fidelity.
Furthermore, we show that single-mode optical cavities outperform bi-modal cavities for both optical initialisation and optical readout across all parameter regimes studied, and can be used to achieve an optimal fidelity in both stages using a single set of cavity parameters.

The paper is organised as follows: in Section~\ref{sec:theory} we introduce the background theory for a charged QD interacting with a single- or bi-modal cavity. 
Sections~\ref{sec:init} and~\ref{sec:read} investigate how the two cavity configurations impact spin initialisation and readout respectively in isolation. 
We then discuss the best cavity configurations for initialisation and readout in tandem in Section~\ref{sec:disc}. 

\section{Background Theory}
\label{sec:theory}

In the absence of any applied magnetic fields, a single electron confined to a QD possesses a spin degree of freedom with two degenerate ground states $\{\ket{\uparrow}_z, \ket{\downarrow}_z\}$ defined along the QD growth ($z$) axis, which are chosen to have zero energy. 
Optically exciting the QD introduces an exciton, forming two negatively charged trion states $\{\ket{\downarrow\uparrow, \Downarrow}_z, \ket{\uparrow\downarrow, \Uparrow}_z\}$ with energy $\hbar\omega_\mathrm{0}$. 
Applying a Voigt geometry magnetic field to the QD lifts the energy degeneracy of both the ground and excited spin states, splitting them by the Zeeman energies $\Delta_B^{e} = g_{e} \mu_B B$ and $\Delta_B^{h} = g_{h} \mu_B B$ respectively (Fig.~\ref{fig:cavity_config}a).
Here $\mu_B$ is the Bohr magneton, $g_{e}$ and $g_{h}$ are the electron and hole effective in-plane g-factors, $B$ is the applied magnetic field strength, and we have ignored any diamagnetic shift.
For convenience, we redefine the QD spin states in the basis along the magnetic field axis as $\{\ket{1}=\ket{\uparrow}_x, \ket{2}=\ket{\downarrow}_x, \ket{3}=\ket{\downarrow\uparrow,\Downarrow}_x, \ket{4}=\ket{\uparrow\downarrow,\Uparrow}_x\}$.
Unless otherwise stated, we assume an applied field of $B=5$~T, which yields $\Delta_\mathrm{B}^\mathrm{e}/2\pi = 35$~GHz and $\Delta_\mathrm{B}^\mathrm{h}/2\pi = 20$~GHz, with $g_\mathrm{e} = 0.5$ and $g_\mathrm{h} = 0.3$ \cite{emary2007fast}.

In the Voigt geometry, there are four allowed optical transitions with equal magnitude,
forming two $\Lambda$-systems coupling each excited state to both ground states via two orthogonal, linearly polarised transitions (Fig.~\ref{fig:cavity_config}a).
While quasi-cycling transitions are most easily introduced in the Voigt geometry via coupling to a cavity with a single, linearly-polarised mode~\cite{PhysRevLett.113.093603,Reinhard2012,doi:10.1063/1.3579535} (Fig.~\ref{fig:cavity_config}b), there are many photonic structures, such as micropillar cavities~\cite{Reithmaier2004}, point-defect photonic crystal (PhC) cavities~\cite{Liu2018}, crossed nanobeam cavities~\cite{Rivoire2011}, and open-access microcavities~\cite{Tomm2021bright}, that naturally possess two orthogonal, linearly-polarised modes (Fig.~\ref{fig:cavity_config}c).
While such bi-modal cavities have been used to suppress resonant laser background in QD single-photon sources~\cite{wang2019towards,Tomm2021bright}, there remain open questions as to the impact on the orthogonally polarised transitions necessary for spin initialisation and readout.
Thus in our set-up each of the $\Lambda$-systems transitions can occur through coupling to the free electromagnetic vacuum, or via an optical cavity. 
The latter is included in the system Hamiltonian $H_{S} = H_0 + H_I,$ 
where ($\hbar=1$)
\begin{align}
    H_0 &= \frac{\Delta_\mathrm{B}^\mathrm{e}}{2}(\sigma_{22} -\sigma_{11}) + \left(\omega_0 - \frac{\Delta_\mathrm{B}^\mathrm{h}}{2}\right)\sigma_{33}\nonumber\\
    &+ \left(\omega_0 + \frac{\Delta_\mathrm{B}^\mathrm{h}}{2}\right)\sigma_{44}
    +\sum_{\lambda=X,Y}\nu_{\lambda}a^\dagger_\lambda a_\lambda.
    \label{eq:H0}
\end{align} 
Here we have defined the spin operators as $\sigma_{ij} = \ket{i}\!\bra{j}$, 
and introduced the cavity mode creation (annihilation) operators $a^\dagger_\lambda$ ($a_\lambda$) with frequency $\nu_\lambda$, where $\lambda=X,~Y,$ denotes the polarisation of the cavity mode.
In physical systems the degeneracy of the bi-modal cavity is often lifted either by intentional design \cite{Coles2014, Gur2021, Gayral1998, Rivoire2011} or by fabrication imperfections~\cite{doi:10.1063/1.2749862,doi:10.1063/1.3696036}, and we therefore assume the cavity modes are detuned.

Applying the rotating wave approximation to our Hamiltonian, the light-matter interaction takes a Jaynes-Cummings form:
\begin{equation}
    H_{I} = \sum_{\lambda=X,Y}g_\lambda a^\dagger_\lambda\bm{\sigma}_\lambda +g^\ast_\lambda a_\lambda\bm{\sigma}^\dagger_\lambda,
    \label{eq:HInt}
\end{equation}
where $g_\lambda$ is the light-matter interaction strength for the relevant cavity mode, and we have introduced the collective transition operators $\bm{\sigma}_X=\sigma_{14} + \sigma_{23}$ and $\bm{\sigma}_Y=\sigma_{24} + \sigma_{13}$ 
(Fig~\ref{fig:cavity_config}c). 
With the above definitions, we can recover the single $Y-$polarised mode cavity set-up depicted in Fig.~\ref{fig:cavity_config}b by setting $g_X=\nu_X=0$.

Spin initialisation and readout necessitate coherent driving of the spin and cavity degrees of freedom respectively.
This is included semi-classically in the model via time-dependent driving terms with frequency $\omega_l$. 
This leads to a total Hamiltonian of the form $H_T(t) = H_0 + H_I + H_D^{QD}(t) + H_D^{C}(t)$, where in the dipole and rotating wave approximations the QD driving term may be written as:
\begin{align}
    H_D^{QD}(t)&= -\frac{1}{2}\sum_{\lambda=X,Y}\Omega_\lambda(t)e^{i\omega_l t}\bm{\sigma}_\lambda +\mathrm{h.c.}.
    \label{eq:HDQD}
\end{align}
Here $\Omega_\lambda(t)$ is the time-dependent Rabi frequency for the relevant polarisation mode.
Direct excitation of the QD occurs when the excitation laser is spatially or spectrally decoupled from the cavity mode(s).
That is when the laser is orientated orthogonally to the cavity axes or far detuned from the cavity resonance.
The cavity driving Hamiltonian takes a similar form to the direct QD excitation Hamiltonian:
\begin{equation}
      H_D^{C}(t) =-\sum_{\lambda=X,Y}\epsilon_\lambda(t)e^{i\omega_l t}a_\lambda +  \mathrm{h.c.},
   \label{eq:HDCav}
\end{equation}
where $\epsilon_\lambda(t)$ is the time-dependent cavity driving strength.
Optical excitation via a cavity mode occurs when the excitation laser is spatially coupled to the mode and spectrally near the cavity resonance.
For convenience, we work in a rotating frame with respect to the laser frequency $\omega_l$.

During spin initialisation the two excitation schemes defined here are qualitatively equivalent, with the only difference being the required excitation power.
Here we only consider optical initialisation via direct excitation to enable a fair comparison between the cavity configurations.
On the other hand, optical readout directly probes the properties of the cavity mode and thus requires cavity driving.

In addition to the unitary dynamics generated by the Hamiltonian $H_\mathrm{T}(t)$, there are also loss processes acting on the cavity-QD system, namely the emission of photons via the cavity or the electromagnetic vacuum. 
These are accounted for through a standard Lindblad master equation of the form~\cite{carmichael1999statistical}:
\begin{align}
    \frac{\partial\rho(t)}{\partial t} = &-i[H_T(t), \rho(t)]\nonumber\\
    &+ \sum_{\lambda=X,Y} \frac{\kappa_\lambda}{2}\mathcal{L}_{a_\lambda}[\rho(t)]+\frac{\gamma_\lambda}{2}\mathcal{L}_{\bm{\sigma}_\lambda}[\rho(t)],
    \label{eq:meq}
\end{align}
where $\rho(t)$ is the reduced density matrix of the cavity-QD system, and $\mathcal{L}_{O}[\rho] = 2 O\rho O^\dagger - \{O^\dagger O,\rho\}$ is the Lindblad superoperator.
Equation~\ref{eq:meq} captures the emission of photons in a given  polarisation state $\lambda$ via two different channels: the first is leakage from the cavity mode, occurring with a rate $\kappa_\lambda$; the second is spontaneous emission directly from the 4-level system (4LS) with rate $\gamma_\lambda$.
Throughout this paper we shall assume that both polarisation transitions have the same lifetime, such that $\gamma_{\lambda=X,Y}^{-1} = \gamma^{-1} = 1$ ns, and in the bi-modal case we assume the cavity modes have identical linewidths ($\kappa_{\lambda=X,Y} = \kappa$) such that $\abs{g_{X,Y}} = \abs{g}$ ($g_X=g$, $g_Y=ig$).
The effects of pure dephasing are considered in Appendix~\ref{AppendixB}, and are shown to be negligible in most cases.
All calculations presented in this paper were performed using the Python package QuTiP~\cite{johansson2012qutip}.

\section{Spin initialisation}\label{sec:init}

\begin{figure*}
    \centering
    \includegraphics[width=17cm]{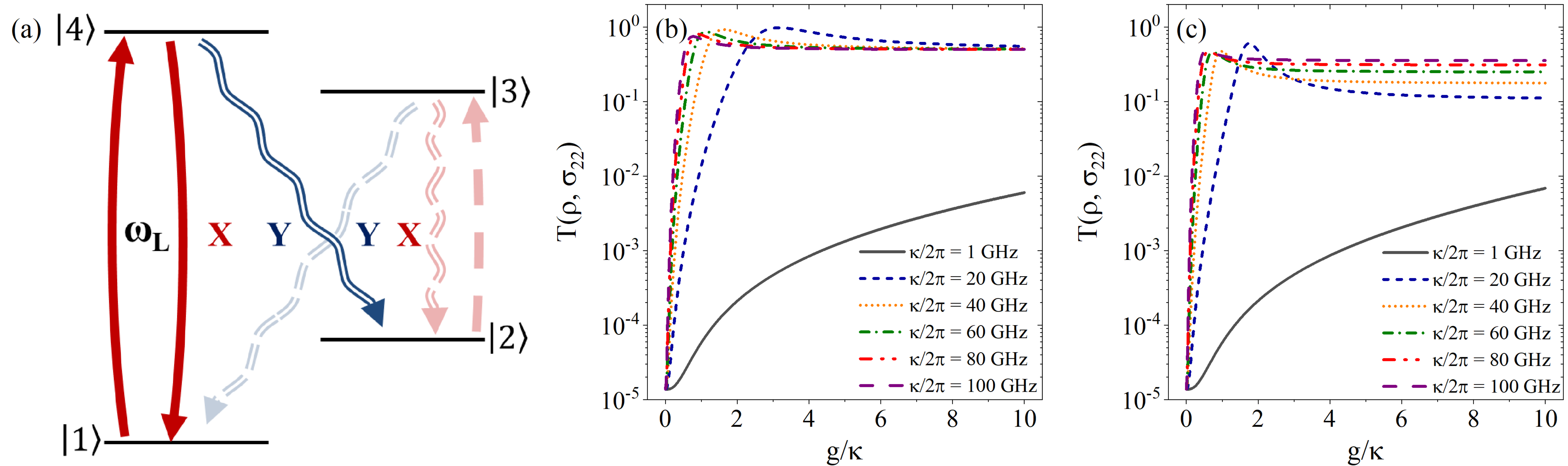}
    \caption{(a) A schematic of the initialisation process. The $\ket{1}\rightarrow\ket{4}$ transition is coherently driven on resonance ($\omega_l = \omega_0 + (\Delta_B^e + \Delta_B^h)/2$) transferring the spin population to $\ket{2}$ via the $\ket{4}\rightarrow\ket{2}$ transition. Off-resonant driving of the $\ket{2}\rightarrow\ket{3}$ transition returns the spin population to $\ket{1}$. (b), (c) The trace distance between the system steady-state after the initialisation process and the $\ket{2}$ ground state as a function of $g/\kappa$ when coupled to a $Y-$polarised single-mode cavity and a bi-modal cavity respectively. Parameters: $B=$ 5 T, $g_{e(h)}=$ 0.5(0.3), $\gamma^{-1}$ = 1 ns, $\Omega_X/2\pi = 10^{-3}$ GHz, $\nu_X = \omega_0$ and $\nu_Y = \omega_0 + (\Delta_B^h - \Delta_B^e)/2$.}
    \label{fig:SteadyState_Init}
\end{figure*}

Assuming an initial state with $\rho_{jj} = 0.5~\mathrm{for}~j\in\{1,2\}$ and $\rho_{ij}=0$ otherwise, we wish to prepare the system in the spin state $\ket{2}$ using the protocol presented in \cite{PhysRevLett.111.027402, Loo2011} and illustrated in Fig.~\ref{fig:SteadyState_Init}a. 
Initialisation is achieved by resonantly driving the  $\ket{1}\rightarrow\ket{4}$ transition using an $X-$polarised laser (i.e. $\Omega_\mathrm{X}>0$, $\Omega_\mathrm{Y}=0$), such that $\omega_l = \omega_0 + (\Delta_B^e + \Delta_B^h)/2$. 
The state $\ket{2}$ is then populated by the $\ket{4}\rightarrow\ket{2}$ transition. 
We choose to initialise the spin state by driving the $X-$polarised transitions owing to the larger detuning between the transitions relative to the laser bandwidth, minimising the off-resonant processes that induce unwanted spin flips away from the $\ket{2}$ target state.

To quantify the initialisation fidelity we use the trace distance, $T(\rho,\varrho) = \frac{1}{2}\Tr{\sqrt{(\rho - \varrho)^2}}$, between the prepared ($\rho$) and target ($\varrho$) states~\cite{Gilchrist2005}. 
The trace distance provides a measure of the distinguishability of two given states using the entirety of the density matrix including all state populations and coherences.
For the pure target state used in this case ($\varrho=\sigma_{22}$) this definition is equivalent to the fidelity used in other studies~\cite{PhysRevLett.126.013602, emary2007fast, Paspalakis2019}. 
Here we choose to quantify the success of the initialisation process using the more general trace distance as this is often simpler to calculate, and provides a true metric on density matrix space, naturally distinguishing between coherent superposition and mixed spin states which would be overlooked if relative spin populations were used as a measure of preparation fidelity \cite{PhysRevLett.111.027402,Gerardot2008, Stefanatos2019, Kumar2016}. 
Using this metric, $T(\rho=\sigma_{22},\sigma_{22}) = 0$ indicates the prepared state is indistinguishable from the target state (unity initialisation fidelity). 
Conversely, $T(\rho,\sigma_{22})=1$ indicates the prepared state is orthogonal to the target state, and is therefore completely distinguishable.

For both the single and bi-modal cavity configurations we choose the $Y$-polarised mode to be resonant with the $\ket{2}\rightarrow\ket{4}$ transition (i.e. $\nu_Y = \omega_{0}+(\Delta_B^h-\Delta_B^e)/2$).
However, when coupled to a bi-modal cavity we leave the $X-$polarised cavity mode detuned from the corresponding transitions such that $\nu_X = \omega_0$ as this has been shown to maximise the initialisation fidelity with this cavity configuration \cite{PhysRevLett.111.027402}.

To allow a direct comparison between the two cavity configurations, we assume spin initialisation is achieved by directly driving the QD transitions.
This differs to previous work studying spin initialisation with bi-modal cavities, which use cavity driving rather than direct QD driving to initialise the spin state~\cite{PhysRevLett.111.027402}.
However, both schemes lead to qualitatively the same behaviour, and importantly considering only one driving configuration allows a fair and consistent comparison between the individual cavity set-ups~\footnote{In the CW driving regime, the two driving regimes are unitarily equivalent.}.

\subsection{Steady-State Limit}

Using the model in Sec.~\ref{sec:theory}, we begin by studying initialisation in the steady-state limit with a continuous wave (CW) driving term (i.e.~$\Omega_{X}(t)~=~\Omega_X~\forall~t$).
While this limit does not accurately reflect experimental procedures for initialising spin states, the steady-state still provides an insight into the behaviour of the system and limits the available parameter space.
Figures~\ref{fig:SteadyState_Init}b and~\ref{fig:SteadyState_Init}c show the calculated trace distance between the prepared steady-state and target state with a fixed driving strength as a function of $g/\kappa$ for a range of cavity linewidths.

The results presented in Fig.~\ref{fig:SteadyState_Init} show that in the steady-state limit the inclusion of cavity effects reduces the initialisation fidelity.
For small $g$ this is a result of the cavity modifying lifetime of the trion states, and thus the ratio $\Omega/\gamma$.
As the cavity coupling strength is increased the fixed driving strength is no longer optimised to achieve the smallest trace distance.
We therefore find that for each set of cavity parameters the Rabi frequency needs to be optimised to minimise the trace distance.
As the cavity-coupled system enters the strong coupling regime ($g\gg\kappa,\gamma$) the QD states hybridise with the cavity modes, fundamentally changing the system eigenstructure, which leads to a maximally mixed ground state, where $T(\rho, \sigma_{11}) = T(\rho, \sigma_{22}) = 0.5$.
For a bi-modal cavity set-up we find the steady-state evolves to return $0.1\leq T(\rho,\sigma_{22})\leq0.5$ depending on the cavity linewidth.

In the limit of the narrowest cavity linewidths, we expect both cavity configurations to display similar behaviour.
At these linewidths only the $\ket{4}\rightarrow\ket{2}$ transition experiences a significant Purcell enhancement in either cavity configuration, with all other optical transitions sufficiently detuned from the cavity mode to experience little to no enhancement.
This expectation is borne out in Fig.~\ref{fig:SteadyState_Init} with $\kappa/2\pi = 1$ GHz.
At this cavity linewidth both cavity configurations return similar trace distances, with small differences resulting from some non-zero enhancement of the $X-$polarised transitions when coupled to a bi-modal cavity.

With large coupling strengths, we find the bi-modal cavity outperforms the single-mode cavity, returning smaller trace distances for $1~\mathrm{GHz}~ < \kappa/2\pi < 100$ GHz.
At intermediate cavity linewidths, the Purcell enhancement of the $\ket{3}\rightarrow\ket{2}$ transition is greater than that of the $\ket{3}\rightarrow\ket{1}$ transition.
Thus any population in the $\ket{3}$ state excited through off-resonant driving will preferentially decay back to the desired $\ket{2}$ state, providing additional protection to the prepared state.
While the Purcell enhancement of the $\ket{4}\rightarrow\ket{1}$ transition does hinder the initialisation process, this effect is less significant in the limit of infinite driving time. 
Increasing the cavity linewidth begins to equalise the enhancement of the transitions away from the $\ket{3}$ state, hence the system tends towards the maximally mixed ground state with $T(\rho,\sigma_{22})=0.5$ as in the single-mode case.

\subsection{Finite Pulse Duration}

\begin{figure*}[t]
    \centering
    \includegraphics[width = 15 cm]{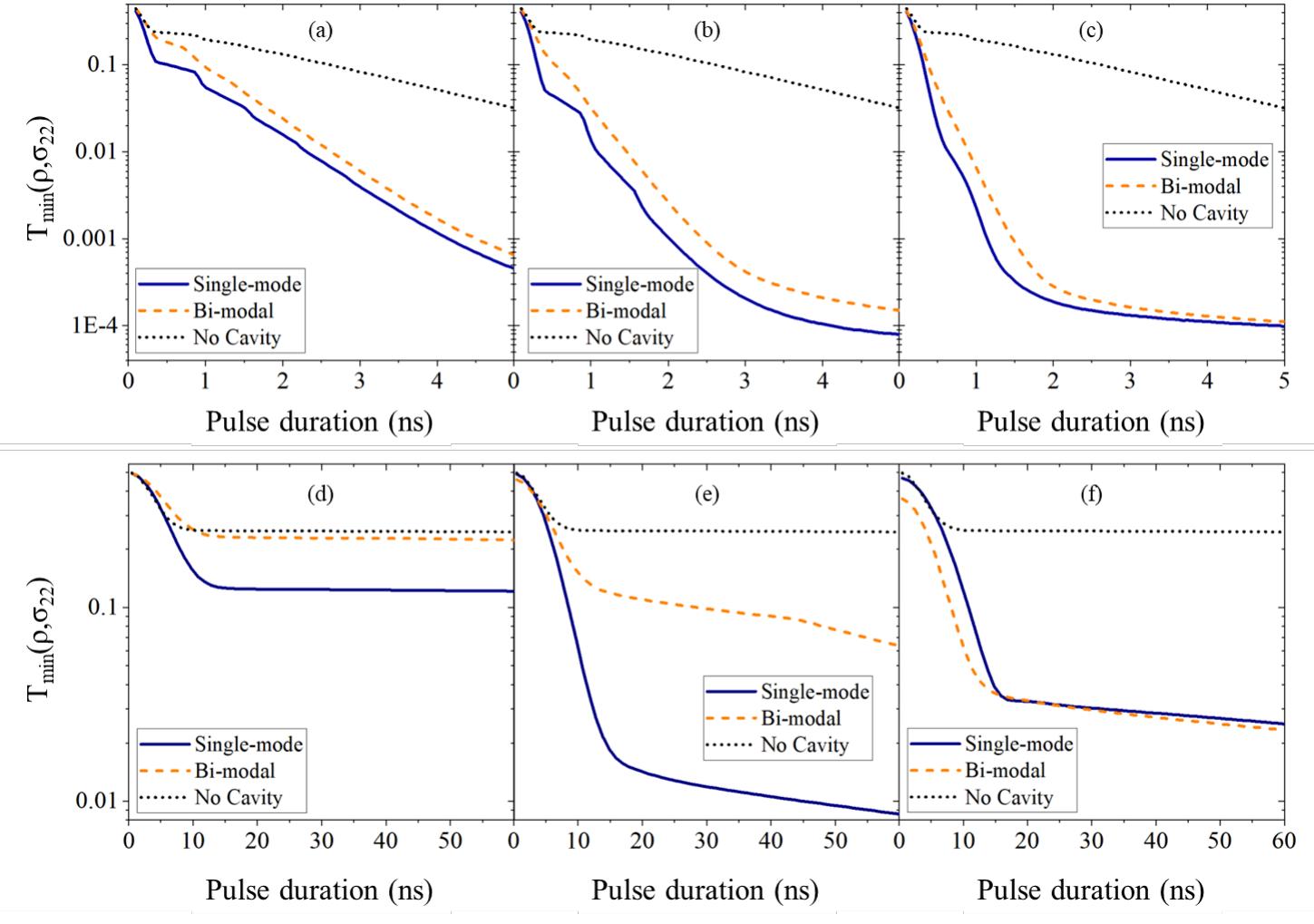}
    \caption{The calculated trace distances between the final state of the four-level system and the target state as a function of the duration of the driving pulse when driving with a (top) top-hat or (bottom) Gaussian pulse. The trace distances are presented for the single-mode (blue solid), and bi-modal (orange dash) cavity configurations as well as without a cavity (black dot), and are minimised with respect to the driving strength (peak Rabi frequency and pulse area for square and Gaussian pulses respectively). The cavity parameters used were: (a) $\kappa/2\pi = 110$ GHz, $F_P = 2$, (b) $\kappa/2\pi = 47
    $ GHz, $F_P = 5$, (c) $\kappa/2\pi = 21$ GHz, $F_P = 10$, 
    (d) $\kappa/2\pi = 110$ GHz, $F_P = 2$, (e) $\kappa/2\pi = 33$ GHz, $F_P = 40$, (f) $\kappa/2\pi = 6$ GHz, $F_P = 40$. Other parameters: $B=$ 5 T, $g_{e(h)}=$ 0.5(0.3), $\gamma^{-1}$ = 1 ns, $\nu_X = \omega_0$, and $\nu_Y = \omega_0 + (\Delta_B^h - \Delta_B^e)/2$.}
    \label{fig:Init_FinitePulse}
\end{figure*}

While the limit of infinite driving time provides some insight into the system behaviours, it does not accurately reflect the experimental realisation of optical spin initialisation.
Any experimentally relevant protocol requires the initialisation process to occur in a finite time, and is thus achieved with finite optical pulses rather than CW drive.
Maximising the spin initialisation fidelity in this limit of finite driving time requires maximising the efficiency and speed with which the spin population is transferred from the $\ket{1}$ state to the $\ket{2}$ state.
Thus we now consider the impact of cavity coupling on the initialisation of the spin system when driven by a finite optical pulse with either a square or Gaussian envelope.

As pulsed optical driving is a more accurate representation of the experimental realisation of optical spin initialisation, we now opt to calculate $T(\varrho,\sigma_{22})$ as a function of more experimentally accessible parameters, namely the Purcell factor and cavity linewidth.
The combination of the light-matter coupling strength and cavity linewidth leads to an enhanced emission rate from the relevant optical transition 
quantified through the Purcell enhancement $F_P(\lambda) = 4\vert g_\lambda\vert^2/\kappa_\lambda\gamma_\lambda$ when on resonance~\cite{Liu2018}.
We shall restrict the cavity linewidth and Purcell factor to $1\leq\kappa/2\pi\leq 110$ GHz and $1\leq F_P\leq 40$ respectively.
This maintains the Purcell enhancement in a regime that has been experimentally demonstrated \cite{Liu2018}, and limits the cavity quality factors to experimentally achievable values (on the order of $10^3-10^5$ for wavelengths in the NIR and telecommunications bands).

\subsubsection{Square Pulse}
We first examine initialisation with a finite square pulse with $\Omega_X(t)= \Omega(H(t-t_0 +\Delta\tau_X/2) - H(t-t_0 -\Delta\tau_X/2))$, where $H(x)$ is the Heaviside function, $t_0$ is the centre of the pulse,  $\Delta\tau_X$ is the pulse duration, and $\Omega$ is the Rabi frequency of the pulse.
For a given set of cavity parameter combinations, we vary the Rabi frequency of the pulse in the range $0\leq\Omega_X\leq 10\gamma$ to find the minimum trace distance for each pulse duration $\Delta\tau_X$, leaving adequate time after the pulse for the trion populations to fully decay. 
Figures~\ref{fig:Init_FinitePulse}a-c show the resulting minimised trace distance as a function of the pulse duration for single-mode and bi-modal cavity structures for different cavity parameters.
The effects of pure dephasing are discussed in Appendix~\ref{AppendixB}, and are shown to have similarly negligible effects for either cavity configuration when driving with a square pulse.

Irrespective of cavity configuration, we find a general trend of decreasing trace distance with increasing pulse duration.
Longer pulse durations both increase the fraction of the $\ket{1}$ state population transferred to the excited state, and, when the pulse duration is longer than the excited state lifetime, enable the re-excitation of any non-zero $\ket{1}$ population resulting from the decay of the trion states via the $\ket{4}\rightarrow\ket{1}$ and $\ket{3}\rightarrow\ket{1}$ transitions.
Furthermore, increasing the duration of the driving pulse decreases its bandwidth which in turn reduces the strength of the off-resonant driving of the $\ket{2}\rightarrow\ket{3}$ transition that moves the system away from the target state.

When coupled to a single-mode cavity and driving with a square pulse, we find the Purcell enhancement of the resonant $Y-$polarised transition is the most important factor in achieving a high initialisation fidelity. 
For the majority of the cavity parameter combinations studied, the effect of the cavity linewidth only becomes significant at longer pulse durations, depending on the Purcell factor.
We find that neither the largest Purcell factor nor the narrowest cavity linewidth studied necessarily produces the best fidelity in the limit of finite pulse duration.
Instead, of the cavity parameter combinations studied, an intermediate linewidth of $\kappa/2\pi = 20$~GHz and Purcell factor of $F_P = 10$ produced the smallest trace distance over the largest range of pulse durations.
Maximising the initialisation fidelity requires balancing the rates at which the system is driven to and from the target state.
An intermediate cavity linewidth is required to avoid the cyclical re-excitation of the $\ket{4}$ state that would occur in the strong coupling regime, and act to increase $T(\varrho,\sigma_{22})$.
However, to limit the rate at which the system is driven away from the target $\ket{2}$ state owing to the increased spectral bandwidth of the finite pulse, the Purcell factor must also be limited to minimise the decay rate of the unwanted $\ket{3}\rightarrow\ket{1}$ transition.

Similar results are found when the QD is coupled to a bi-modal cavity. 
Again the Purcell enhancement of the resonant $Y-$polarised transition appears to be the dominant factor in determining the trace distance as a function of pulse duration. 
However, relative to initialisation with a single-mode cavity, the calculated trace distances are larger and require a longer pulse duration to achieve. 
In this configuration, we find a Purcell factor of $F_P=40$ produces the largest trace distances for the majority of cavity linewidths studied.
For the cavity parameters considered, we find that $F_P=10$ and $\kappa/2\pi=1$~GHz produces the best initialisation fidelity with pulse durations less than $\approx 0.5$~ns while $F_P=10$ and $\kappa/2\pi=20$~GHz produces the best initialisation fidelity for pulse durations greater than $\approx 0.5$~ns.

Plateaus and regions of decreased gradient in the trace distance can also be seen in Figs.~\ref{fig:Init_FinitePulse}a-c. 
These plateaus are artefacts of limiting range over which the Rabi frequency is swept for each data point.
Driving the system with increasing Rabi frequency naturally results in Rabi oscillations in the system populations.
The edge of each plateau occurs when the pulse duration is long enough to encompass the next oscillation with a lower local minimum than the previous. 
These plateaus are not seen at longer pulse durations as the global minimum in the trace distance as a function of Rabi frequency usually occurs after two or three oscillations.

\subsubsection{Gaussian Pulse}

We now move on to study initialisation with a Gaussian optical pulse.
We define the Gaussian pulse with polarisation $\lambda=X,Y$ centered around $t_0$ as:
\begin{equation}
    \Omega_\lambda(t) = \frac{\Theta_\lambda}{\sqrt{2\pi w_\lambda^2}}\exp\left\{{-\frac{\left(t - t_0\right)^2}{2w_\lambda^2}}\right\},
\end{equation}
where $\Theta_\lambda=\int_{-\infty}^\infty dt\Omega_\lambda(t)$ is the pulse area defined such that a pulse with $\Theta_\lambda=\pi$ would invert the population of a two-level system, and $w_\lambda$ is the Gaussian width of the pulse related to the intensity Full-Width Half-Maximum ($\Delta\tau_\lambda$) by:
\begin{equation}
    w_\lambda = \frac{\Delta\tau_\lambda}{2\sqrt{\ln{2}}}.
\end{equation}

Following the procedure for the finite square pulse, we plot the trace distance, minimised with respect to the pulse area in the range $0.01\pi\leq\Theta_X\leq5\pi$, as a function of $\Delta\tau_X$.
Figures~\ref{fig:Init_FinitePulse}d-f show the results for Gaussian pulsed excitation.

Just as with the square optical pulse, we find the trace distance decreases with increasing pulse duration.
Spin initialisation with Gaussian pulses occurs on much shorter timescales than when using square pulses, with the trace distance being optimised in tens of picoseconds rather than a few nanoseconds. 
Additionally, the trace distances achieved with a Gaussian pulse are smaller than those achieved with the shortest square pulses studied ($<1$ ns) which are unlikely to be experimentally accessible.
However, the smallest achieved trace distances driving with a Gaussian pulse are orders of magnitude larger than those found when driving with a longer square pulse ($>1$ ns).
Below $\Delta\tau_X\lessapprox 15$ ps this is a consequence of the bandwidth of the driving pulse resulting in a significant spectral overlap between the driving field and the off-resonant $\ket{2}\rightarrow\ket{3}$ transition, which in turn increases the driving of the system away from the desired state.
However, beyond $\Delta\tau_X \approx 15$ ps this overlap is minimised, and thus the trace distance is governed by re-excitation of population that initially decayed into the unwanted ground state ($\ket{1}$).
This in turn is limited by the ratio of the decay rate of the trion states into the $\ket{1}$ state and the pulse duration.
Furthermore, attaining these trace distances when driving with a Gaussian pulse requires the largest Purcell enhancement for all but the narrowest cavity linewidths due to the pulse duration being much shorter than the trion lifetimes, which can prove challenging to realise experimentally.

We again find coupling the QD to a bi-modal cavity significantly increases the calculated trace distances after the initialisation procedure.
In fact, for some of the cavity parameters studied for the bi-modal case, the produced trace distances are comparable to initialisation with a Gaussian pulse in the absence of any cavity effects.
However, Appendix~\ref{AppendixB} shows that when driving with a Gaussian pulse, spin initialisation with a bi-modal cavity configuration is the most robust against pure dephasing.

\section{Spin Readout}
\label{sec:read}
\begin{figure*}[ht!]
    \centering
    \includegraphics[width=17.5cm]{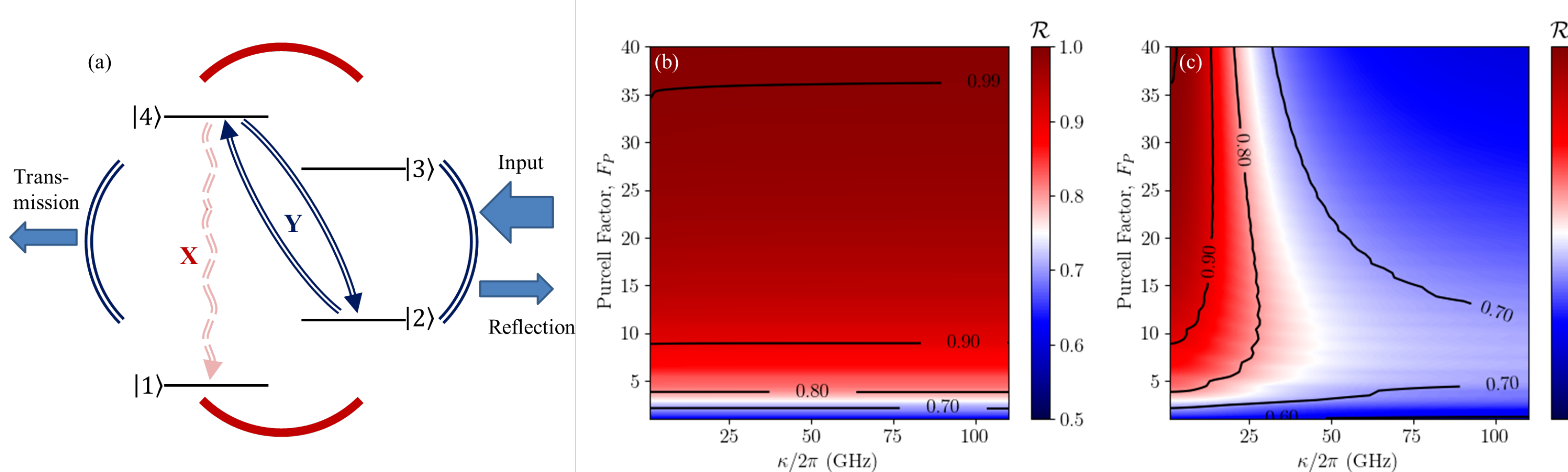}
    \caption{(a) A diagram of the spin-readout process. The $Y$-polarised cavity mode is probed with a square laser pulse and the QD-state dependent reflectivity/transmissivity is measured. By comparing the collected photon number with a threshold value the spin state of the system may be determined. (b) and (c) The calculated readout fidelities ($R$) for a single mode cavity, and a bi-modal cavity respectively, when driving the $Y-$polarised cavity mode with a 35 ns square pulse. Parameters: $B=$ 5 T, $g_{e(h)}=$ 0.5(0.3), $\gamma^{-1}$ = 1 ns, $\epsilon_Y = \sqrt{(0.01\times 2g_Y^2)}$, $\eta = 1$, $\nu_X = \omega_0$ and $\nu_Y = \omega_0 + (\Delta_B^h - \Delta_B^e)/2$.}
    \label{fig:Readout_SuccessProb}
\end{figure*}

In addition to high fidelity spin initialisation, it is crucial that the cavity structures are also conducive to high fidelity readout. 
We therefore study optical spin readout for both single mode and bi-modal cavities using the method first proposed in \cite{PhysRevA.94.012307}, and demonstrated experimentally in~\cite{PhysRevApplied.9.054013}. 
This readout method uses the spin-dependent transmissivity or reflectivity of a cavity mode resonantly coupled to a transition of the QD to determine its spin state.
In contrast to Ref.~\cite{PhysRevA.94.012307}, we choose the cavity configuration to mirror the set-up used for high-fidelity spin initialisation, where the cavity mode is coupled to the $Y-$polarised transitions.
The $\ket{2}\rightarrow\ket{4}$ transition is then weakly probed by a resonant drive via the $Y-$polarised cavity mode present in either cavity configuration.
If the emitter occupies the $\ket{1}$ state the photons are transmitted, while if the QD resides in the $\ket{2}$ state the photons are reflected (see Fig.~\ref{fig:Readout_SuccessProb}a).
By comparing the ratio of photons transmitted versus incident with a threshold value \cite{PhysRevA.94.012307}, the state ($\ket{1}$ or $\ket{2}$) occupied by the QD can be determined.

The readout protocol is initialised by weakly probing the $Y$-polarised cavity mode over some time-interval $[0,\tau]$, where $\tau$ is much longer than the lifetime of the system.
Over this time interval, we set a threshold photon number $k$, where if the number of collected transmitted photons is less than $k$ then the qubit state is $\ket{2}$, otherwise it is $\ket{1}$.
The maximum probability of a successful readout occurring is then given by $\mathcal{R} = \max_k(q_1 p_1(k) + q_2 p_2(k))$, where $q_i$ is the probability of finding the qubit in state $i$, and $p_i$ is the probability of getting a correct result using threshold photon number $k$.
As shown by Ref.~\cite{PhysRevA.94.012307}, for detectors with a dead time shorter than the interval between detection events, and in the weak driving limit, the probabilities $p_i(k)$ can be described by Poissonian statistics. 
This allows the readout fidelity ($\mathcal{R}$) to be written as \cite{PhysRevA.94.012307}:
\begin{multline}
    \mathcal{R}(\tau) = \frac{1}{2} - \frac{1}{2}\sum_{k=0}^M\frac{1}{k!}\Bigl(\left[N_1(\tau)\right]^ke^{-N_1(\tau)} \\ - \left[N_2(\tau)\right]^ke^{-N_2(\tau)}\Bigr),
\end{multline}
where $M$ is the optimal threshold value \cite{PhysRevA.94.012307}:
\begin{equation}
    M = \left\lfloor\frac{N_2(\tau) - N_1(\tau)}{\ln\left[N_2(\tau)\right] - \ln\left[N_1(\tau)\right]}\right\rfloor,
\end{equation}
with $\lfloor x\rfloor$ indicating the largest integer smaller than x, and $N_{i}$ is the number of photons emitted from the cavity mode when the QD starts in the ground state $\ket{i=1,2}$. 
The number of photons emitted can be found by integrating the output flux of the cavity over the duration of the optical readout pulse:
\begin{equation}
    N_{i,\lambda}(\tau) = \eta\kappa_\lambda\int_0^\tau \left|\Tr\left[a^\dagger_\lambda a_\lambda\rho_{i}(t)\right]\right|dt.
\end{equation}
Here $\tau$ is the duration of the readout pulse, $\eta$ is the photon collection efficiency, and $\rho_{i}(t)$ are the density matrices of the system at time $t$ when the QD is initialised in the ground state $\ket{i=1,2}$.
We also continue to assume both modes of the bi-modal cavity have identical linewidths such that $\kappa_{\lambda=X,Y}=\kappa$.
This readout method has been demonstrated to work experimentally, producing $\mathcal{R} = 0.61$ with a cavity linewidth $\kappa/2\pi = 67$ GHz, an enhancement factor $F_P = 62$, a collection efficiency $\eta = 4.1\times 10^{-3}$, and 75 ns pulse duration \cite{PhysRevApplied.9.054013}.

Considering a square pulse with a duration of $35$~ns \footnote{We use a 35 ns readout pulse as a compromise to maximise the readout fidelity across a range of cavity parameters.}, and time-dependent cavity driving strength $\epsilon_Y(t) = \sqrt{(0.01\times 2g_Y^2)}$ for $t\in[0, 35~\mathrm{ns}]$ to remain in the weak-excitation regime \cite{PhysRevA.94.012307}, and assuming $\eta = 1$, we calculate the spin readout fidelity for a range of cavity parameters. 
The weak-excitation regime is used in the readout stage to limit readout induced errors resulting from the readout pulse influencing the state of the system.
Fig.~\ref{fig:Readout_SuccessProb}b and Fig.~\ref{fig:Readout_SuccessProb}c show these results for a single-mode and bi-modal cavity respectively. 
In the case of the single-mode cavity, we find the readout fidelity is primarily dependent on the Purcell enhancement of the $Y-$polarised transitions, and varies little with respect to $\kappa_Y$. 
This is because the Purcell enhancement increases the strength of the quasi-cycling transition required to produce a detectable readout signal, while the cavity linewidth has little impact on the Purcell enhancement of the resonant transition being probed. 
Figure~\ref{fig:Readout_SuccessProb}b shows $F_P = 7$ gives $\mathcal{R}\approx 90\%$.

In contrast, the bi-modal readout fidelity is sensitive to both the cavity linewidth and Purcell enhancement of the optical transitions.
The highest readout fidelities with this cavity configuration are produced with narrow cavity linewidths and large Purcell factors, which prevent undesirable transitions from being Purcell enhanced.
This increases the cyclicity of the $\Lambda$-systems, which naturally increases the strength of the quasi-cycling transition probed during the readout procedure.
To achieve $\mathcal{R}>90\%$ with a bi-modal cavity requires $\kappa/2\pi\leq 13$ GHz and $F_P\geq 9$. 
Note that under these driving conditions, when the cavity parameters are optimised for either cavity configuration, we find that a collection efficiency of $\eta \geq 48\%$ gives $\mathcal{R}\geq 99\%$.
Such collection efficiencies have already been demonstrated in an open-access microcavity system \cite{Tomm2021bright}; for planar PhC or nanobeam structures very high efficiencies could be achieved by direct fibre coupling \cite{Tiecke2015, Daveau:17}.
The effects of pure dephasing on the readout fidelity are shown in Appendix~\ref{AppendixB}.
We find that, while increasing the pure dephasing rate increases the required Purcell factors for both cavity configurations, the single-mode cavity is more robust against these processes in the readout stage.

\section{Conclusion}
\label{sec:disc}

By performing full cavity QED calculations, we have investigated optical spin initialisation and readout for a QD interacting with either single- or bi-modal optical cavities.
Interestingly, we find that for both initialisation and readout, a single mode optical cavity outperforms the bi-modal cavity over the full parameter regime studied regardless of the pulse envelope used.
This is a consequence of the bi-modal cavity Purcell enhancing undesirable transitions, therefore suppressing desired spin-flip processes.
While these unwanted transitions can be suppressed for spin preparation, we find that they restrict the parameter regimes for which high fidelity optical spin readout can be achieved with bi-modal cavities.
Furthermore, this range of optimal readout cavity parameters (large $F_P$) does not overlap with the small range of cavity parameters required for optimised spin initialisation in a bi-modal cavity (small $F_P$).

In contrast, we have shown a near unity readout fidelity is possible with a single-mode cavity across the vast majority of the cavity parameters studied. 
In the single mode configuration with experimentally achievable~\cite{Volz2012,PhysRevB.97.235448,doi:10.1063/1.5144959} cavity parameters ($\kappa/2\pi = 20$ GHz, $F_P = 10$), we find an initialisation trace distance $T(\rho,\sigma_{22})=1.3\times10^{-4}$ is achievable with a 3 ns square optical pulse,
whilst a readout success probability of $\mathcal{R}>0.90$ is possible with a 35 ns optical pulse.
Increasing the Purcell factor to $F_P=35$ increases the readout success rate to $\mathcal{R}=0.99$ while maintaining an initialisation trace distance on the order of $10^{-4}$.
For a bi-modal system using the same parameters, a comparable initialisation trace distance of $T(\rho,\sigma_{22}) = 2.2\times10^{-4}$ is achievable.
However, the corresponding readout success rate is reduced to $\mathcal{R} \approx 0.88$.
In the parameter regimes studied here, no single set of bi-modal cavity parameters simultaneously allow for high fidelity spin initialisation and readout with finite optical pulses.

These results suggest that to realise a high fidelity spin-photon interface, a single linearly-polarised cavity mode providing a modest ($\sim 20$) Purcell enhancement is the optimal configuration.
In addition, further calculations presented in Appendix~\ref{AppendixB} demonstrate that this parameter regime is also robust against significant levels of pure dephasing, illustrating the potential to achieve high performance spin-photon interfaces in real physical systems.
For inherently bi-modal cavities such as micropillars or point-defects in PhCs, our analysis suggests that there would be significant benefit in modifying designs to induce a mode splitting much larger than the cavity linewidth (e.g. elliptical micropillars~\cite{wang2019towards}). 
We believe that these insights and methods will contribute to the development of high fidelity spin-photon interfaces that meet the stringent requirements of future optical quantum technologies.

\section{Acknowledgements}
The authors would like to acknowledge the funding support of the University of Sheffield, and the Engineering and Physical
Sciences Research Council (EPSRC) (UK) grants EP/N031776/1 and EP/V026496/1. AJB acknowledges support from the EPSRC (UK) fellowship EP/W027909/1.
For the purpose of open access, the authors have applied a Creative Commons Attribution (CC BY) license to any Author Accepted Manuscript version arising.

\appendix

\section{Pure Dephasing}\label{AppendixB}

Thus far we have assumed that the trion states are lifetime limited.
However, in physical systems elastic processes will occur that preserve the spin populations, but reduce their coherence.
We therefore also study the effect of pure dephasing of the trion states on the spin initialisation and readout processes when driving with a finite duration pulse.
We account for pure dephasing through the addition of further Lindblad terms,
\begin{equation}
    \sum_{j=3,4}\frac{\Gamma}{2}\mathcal{L}_{\bm{\sigma}_{jj}}[\rho(t)],
\end{equation}
in the master equation.
Here $\Gamma$ is the pure dephasing rate, and $\mathcal{L}_{O}[\rho] = 2 O\rho O^\dagger - \{O^\dagger O,\rho\}$ is the Lindblad super-operator.

\begin{figure}[h]
    \centering
    \includegraphics[width = 8.5cm]{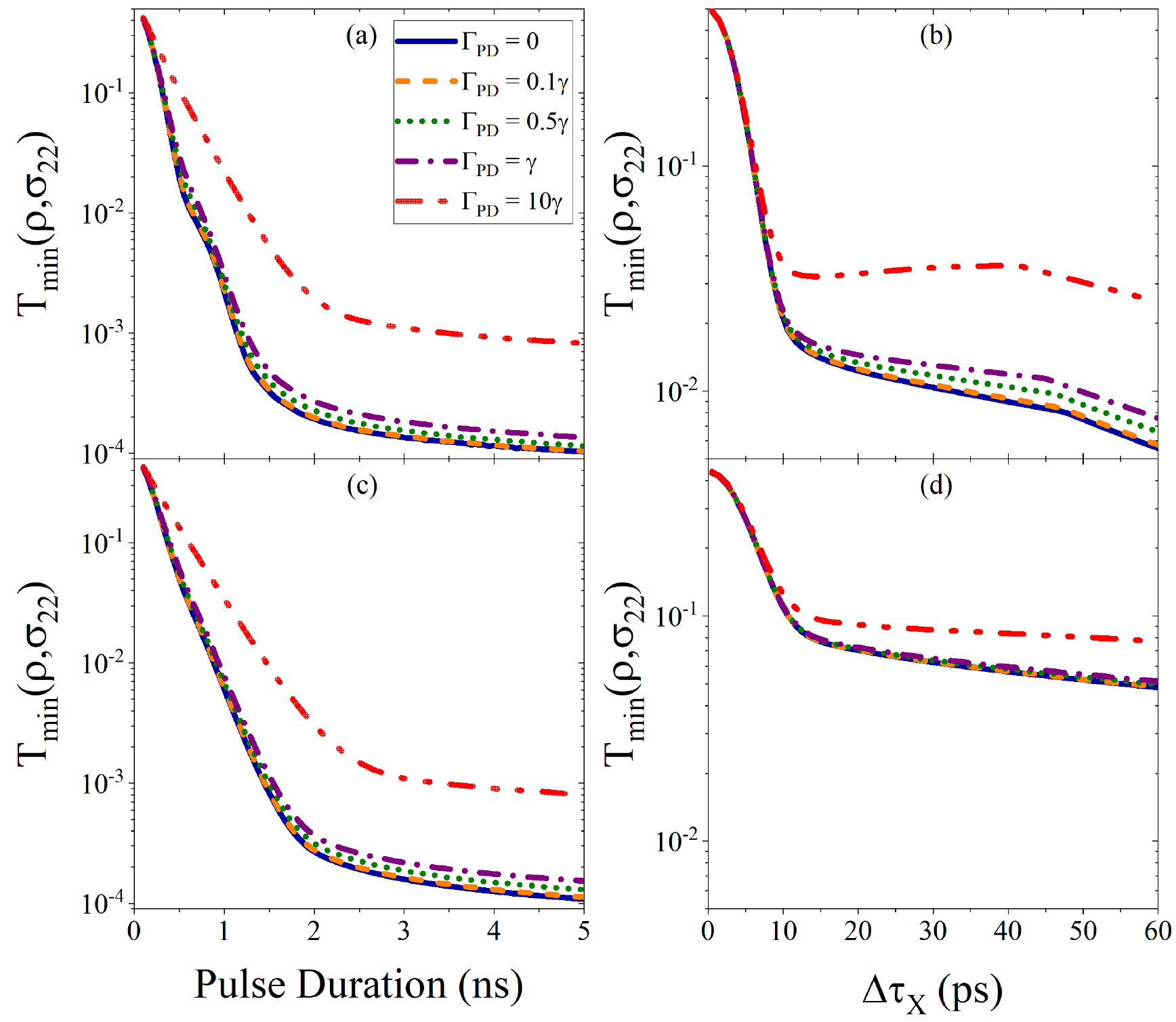}
    \caption{The calculated trace distance between the state prepared by the initialisation process and the target state minimising with respect to the driving strength when driving with a (left: a,c) Square pulse or (right: b,d) Gaussian pulse, and coupled to a (top: a,b) single-mode or (bottom: c,d) bi-modal cavity for different pure dephasing rates indicated in the legend in (a) using the cavity parameters that minimise the trace distance in the absence of pure dephasing processes. Parameters used: $\gamma^{-1}$ = 1 ns, (a,c) $\kappa/2\pi = 20$ GHz, $F_P = 10$, (b) $\kappa/2\pi = 40$ GHz, $F_P = 40$, (d) $\kappa/2\pi = 20$ GHz, $F_P = 40$.}
    \label{fig:Init_PureDeph}
\end{figure}

\begin{figure}[t]
    \centering
    \includegraphics[width=8.5cm]{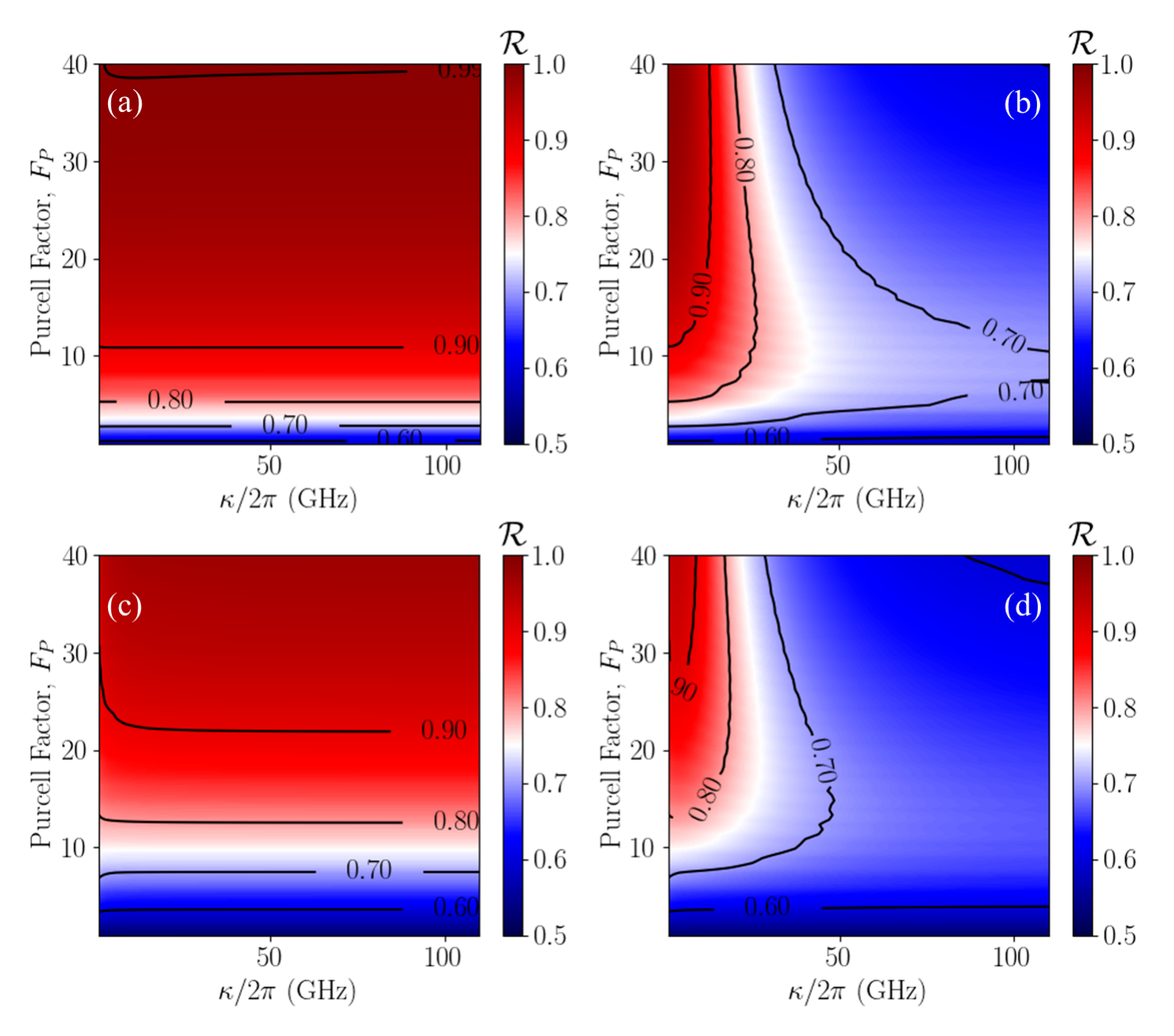}
    \caption{The calculated spin readout success probabilities when coupled to a (left: a,c) single-mode or (right: b,d) bi-modal cavity when the 4LS experiences pure dephasing with rates (top: a,b) $\Gamma = \gamma$ and (bottom: c,d) $\Gamma = 10\gamma$. Parameters used: $B=$ 5 T, $g_{e(h)}=$ 0.5(0.3), $\gamma^{-1}$ = 1 ns, $\epsilon_Y = \sqrt{(0.01\times 2g_Y^2)}$, $\eta = 1$, $\nu_X = \omega_0$ and $\nu_Y = \omega_0 + (\Delta_B^h - \Delta_B^e)/2$.}
    \label{fig:readout_puredeph}
\end{figure}

\subsection{Spin Initialisation}

To study the influence of pure dephasing in the optical initialisation process we follow the same procedure presented in Sec.~\ref{sec:init}, sweeping the duration of the given optical pulse and minimising the trace distance with respect to the Rabi frequency or pulse area.
However, we now only consider the combination of cavity parameters that produced the smallest trace distance in the absence of pure dephasing for a given pulse envelope, and instead plot the calculated trace distances for a number of pure dephasing rates.

Figure~\ref{fig:Init_PureDeph} shows that when $\Gamma\leq\gamma$ pure dephasing has a minimal impact on the calculated trace distances after the initialisation process for any combination of driving pulse envelope and cavity configuration.
For $\Gamma\gg\gamma$, Fig.~\ref{fig:Init_PureDeph} shows pure dephasing significantly increases the calculated trace distances.
For all combinations studied the divergence between the traces distances calculated neglecting and including pure dephasing increases with increasing pulse duration.
However, while producing the largest minimised trace distance, optical spin initialisation with the combination of a bi-modal cavity and Gaussian pulse drive appears the least susceptible to the influence of pure dephasing.
The Purcell enhancement of all four optical transitions, combined with the fast optical driving, work to reduce the relative impact of pure dephasing mechanisms by minimising the time the system spends in the excited states.
Nevertheless, we note that for all configurations, small trace distances are achievable even in the presence of significant pure dephasing.

\subsection{Spin Readout}

To study the impact of pure dephasing on the spin readout process we calculate the readout success probability for two pure dephasing rates, $\Gamma=\gamma$ and $\Gamma = 10\gamma$.
Figure~\ref{fig:readout_puredeph} shows the results when coupled to a single-mode or bi-modal cavity.
When coupled to a single-mode cavity, Figs.~\ref{fig:readout_puredeph}(a) and (c) show that the readout process is impacted by pure dephasing.
While at $\Gamma=\gamma$ $\mathcal{R}>0.99$ is still just achievable in the parameter regimes studied, increasing the pure dephasing rate to $\Gamma=10\gamma$ this is no longer achievable although there remains a significant region where $\mathcal{R}>0.9$.
When coupled to a bi-modal cavity, increasing the pure dephasing rate from $\Gamma=\gamma$ to $\Gamma=10\gamma$ roughly halves the parameter regime where $\mathcal{R}>=0.9$ is achievable.

\bibliography{references}

%apsrev4-2.bst 2019-01-14 (MD) hand-edited version of apsrev4-1.bst
%Control: key (0)
%Control: author (8) initials jnrlst
%Control: editor formatted (1) identically to author
%Control: production of article title (0) allowed
%Control: page (0) single
%Control: year (1) truncated
%Control: production of eprint (0) enabled
\begin{thebibliography}{55}%
\makeatletter
\providecommand \@ifxundefined [1]{%
 \@ifx{#1\undefined}
}%
\providecommand \@ifnum [1]{%
 \ifnum #1\expandafter \@firstoftwo
 \else \expandafter \@secondoftwo
 \fi
}%
\providecommand \@ifx [1]{%
 \ifx #1\expandafter \@firstoftwo
 \else \expandafter \@secondoftwo
 \fi
}%
\providecommand \natexlab [1]{#1}%
\providecommand \enquote  [1]{``#1''}%
\providecommand \bibnamefont  [1]{#1}%
\providecommand \bibfnamefont [1]{#1}%
\providecommand \citenamefont [1]{#1}%
\providecommand \href@noop [0]{\@secondoftwo}%
\providecommand \href [0]{\begingroup \@sanitize@url \@href}%
\providecommand \@href[1]{\@@startlink{#1}\@@href}%
\providecommand \@@href[1]{\endgroup#1\@@endlink}%
\providecommand \@sanitize@url [0]{\catcode `\\12\catcode `\$12\catcode
  `\&12\catcode `\#12\catcode `\^12\catcode `\_12\catcode `\%12\relax}%
\providecommand \@@startlink[1]{}%
\providecommand \@@endlink[0]{}%
\providecommand \url  [0]{\begingroup\@sanitize@url \@url }%
\providecommand \@url [1]{\endgroup\@href {#1}{\urlprefix }}%
\providecommand \urlprefix  [0]{URL }%
\providecommand \Eprint [0]{\href }%
\providecommand \doibase [0]{https://doi.org/}%
\providecommand \selectlanguage [0]{\@gobble}%
\providecommand \bibinfo  [0]{\@secondoftwo}%
\providecommand \bibfield  [0]{\@secondoftwo}%
\providecommand \translation [1]{[#1]}%
\providecommand \BibitemOpen [0]{}%
\providecommand \bibitemStop [0]{}%
\providecommand \bibitemNoStop [0]{.\EOS\space}%
\providecommand \EOS [0]{\spacefactor3000\relax}%
\providecommand \BibitemShut  [1]{\csname bibitem#1\endcsname}%
\let\auto@bib@innerbib\@empty
%</preamble>
\bibitem [{\citenamefont {Kroutvar}\ \emph {et~al.}(2004)\citenamefont
  {Kroutvar}, \citenamefont {Ducommun}, \citenamefont {Heiss}, \citenamefont
  {Bichler}, \citenamefont {Schuh}, \citenamefont {Abstreiter},\ and\
  \citenamefont {Finley}}]{kroutvar2004}%
  \BibitemOpen
  \bibfield  {author} {\bibinfo {author} {\bibfnamefont {M.}~\bibnamefont
  {Kroutvar}}, \bibinfo {author} {\bibfnamefont {Y.}~\bibnamefont {Ducommun}},
  \bibinfo {author} {\bibfnamefont {D.}~\bibnamefont {Heiss}}, \bibinfo
  {author} {\bibfnamefont {M.}~\bibnamefont {Bichler}}, \bibinfo {author}
  {\bibfnamefont {D.}~\bibnamefont {Schuh}}, \bibinfo {author} {\bibfnamefont
  {G.}~\bibnamefont {Abstreiter}},\ and\ \bibinfo {author} {\bibfnamefont
  {J.~J.}\ \bibnamefont {Finley}},\ }\bibfield  {title} {\bibinfo {title}
  {Optically programmable electron spin memory using semiconductor quantum
  dots},\ }\href {https://doi.org/10.1038/nature03008} {\bibfield  {journal}
  {\bibinfo  {journal} {Nature}\ }\textbf {\bibinfo {volume} {432}},\ \bibinfo
  {pages} {81} (\bibinfo {year} {2004})}\BibitemShut {NoStop}%
\bibitem [{\citenamefont {Gillard}\ \emph {et~al.}(2021)\citenamefont
  {Gillard}, \citenamefont {Griffiths}, \citenamefont {Ragunathan},
  \citenamefont {Ulhaq}, \citenamefont {McEwan}, \citenamefont {Clarke},\ and\
  \citenamefont {Chekhovich}}]{gillard2021fundamental}%
  \BibitemOpen
  \bibfield  {author} {\bibinfo {author} {\bibfnamefont {G.}~\bibnamefont
  {Gillard}}, \bibinfo {author} {\bibfnamefont {I.~M.}\ \bibnamefont
  {Griffiths}}, \bibinfo {author} {\bibfnamefont {G.}~\bibnamefont
  {Ragunathan}}, \bibinfo {author} {\bibfnamefont {A.}~\bibnamefont {Ulhaq}},
  \bibinfo {author} {\bibfnamefont {C.}~\bibnamefont {McEwan}}, \bibinfo
  {author} {\bibfnamefont {E.}~\bibnamefont {Clarke}},\ and\ \bibinfo {author}
  {\bibfnamefont {E.~A.}\ \bibnamefont {Chekhovich}},\ }\bibfield  {title}
  {\bibinfo {title} {Fundamental limits of electron and nuclear spin qubit
  lifetimes in an isolated self-assembled quantum dot},\ }\href
  {https://www.nature.com/articles/s41534-021-00378-2} {\bibfield  {journal}
  {\bibinfo  {journal} {NJP Quantum Information}\ }\textbf {\bibinfo {volume}
  {7}},\ \bibinfo {pages} {1} (\bibinfo {year} {2021})}\BibitemShut {NoStop}%
\bibitem [{\citenamefont {Press}\ \emph {et~al.}(2010)\citenamefont {Press},
  \citenamefont {De~Greve}, \citenamefont {McMahon}, \citenamefont {Ladd},
  \citenamefont {Friess}, \citenamefont {Schneider}, \citenamefont {Kamp},
  \citenamefont {Höfling}, \citenamefont {Forchel},\ and\ \citenamefont
  {Yamamoto}}]{Press2010}%
  \BibitemOpen
  \bibfield  {author} {\bibinfo {author} {\bibfnamefont {D.}~\bibnamefont
  {Press}}, \bibinfo {author} {\bibfnamefont {K.}~\bibnamefont {De~Greve}},
  \bibinfo {author} {\bibfnamefont {P.~L.}\ \bibnamefont {McMahon}}, \bibinfo
  {author} {\bibfnamefont {T.~D.}\ \bibnamefont {Ladd}}, \bibinfo {author}
  {\bibfnamefont {B.}~\bibnamefont {Friess}}, \bibinfo {author} {\bibfnamefont
  {C.}~\bibnamefont {Schneider}}, \bibinfo {author} {\bibfnamefont
  {M.}~\bibnamefont {Kamp}}, \bibinfo {author} {\bibfnamefont {S.}~\bibnamefont
  {Höfling}}, \bibinfo {author} {\bibfnamefont {A.}~\bibnamefont {Forchel}},\
  and\ \bibinfo {author} {\bibfnamefont {Y.}~\bibnamefont {Yamamoto}},\
  }\bibfield  {title} {\bibinfo {title} {Ultrafast optical spin echo in a
  single quantum dot},\ }\href {https://doi.org/10.1038/nphoton.2010.83}
  {\bibfield  {journal} {\bibinfo  {journal} {Nature Photonics}\ }\textbf
  {\bibinfo {volume} {4}},\ \bibinfo {pages} {367} (\bibinfo {year}
  {2010})}\BibitemShut {NoStop}%
\bibitem [{\citenamefont {Bechtold}\ \emph {et~al.}(2015)\citenamefont
  {Bechtold}, \citenamefont {Rauch}, \citenamefont {Li}, \citenamefont
  {Simmet}, \citenamefont {Ardelt}, \citenamefont {Regler}, \citenamefont
  {Müller}, \citenamefont {Sinitsyn},\ and\ \citenamefont
  {Finley}}]{Bechtold2015}%
  \BibitemOpen
  \bibfield  {author} {\bibinfo {author} {\bibfnamefont {A.}~\bibnamefont
  {Bechtold}}, \bibinfo {author} {\bibfnamefont {D.}~\bibnamefont {Rauch}},
  \bibinfo {author} {\bibfnamefont {F.}~\bibnamefont {Li}}, \bibinfo {author}
  {\bibfnamefont {T.}~\bibnamefont {Simmet}}, \bibinfo {author} {\bibfnamefont
  {P.-L.}\ \bibnamefont {Ardelt}}, \bibinfo {author} {\bibfnamefont
  {A.}~\bibnamefont {Regler}}, \bibinfo {author} {\bibfnamefont
  {K.}~\bibnamefont {Müller}}, \bibinfo {author} {\bibfnamefont {N.~A.}\
  \bibnamefont {Sinitsyn}},\ and\ \bibinfo {author} {\bibfnamefont {J.~J.}\
  \bibnamefont {Finley}},\ }\bibfield  {title} {\bibinfo {title} {Three-stage
  decoherence dynamics of an electron spin qubit in an optically active quantum
  dot},\ }\href {https://doi.org/10.1038/nphys3470} {\bibfield  {journal}
  {\bibinfo  {journal} {Nature Physics}\ }\textbf {\bibinfo {volume} {11}},\
  \bibinfo {pages} {1005} (\bibinfo {year} {2015})}\BibitemShut {NoStop}%
\bibitem [{\citenamefont {Kimble}(2008)}]{kimble2008quantum}%
  \BibitemOpen
  \bibfield  {author} {\bibinfo {author} {\bibfnamefont {H.~J.}\ \bibnamefont
  {Kimble}},\ }\bibfield  {title} {\bibinfo {title} {The quantum internet},\
  }\href@noop {} {\bibfield  {journal} {\bibinfo  {journal} {Nature}\ }\textbf
  {\bibinfo {volume} {453}},\ \bibinfo {pages} {1023} (\bibinfo {year}
  {2008})}\BibitemShut {NoStop}%
\bibitem [{\citenamefont {Lindner}\ and\ \citenamefont
  {Rudolph}(2009)}]{PhysRevLett.103.113602}%
  \BibitemOpen
  \bibfield  {author} {\bibinfo {author} {\bibfnamefont {N.~H.}\ \bibnamefont
  {Lindner}}\ and\ \bibinfo {author} {\bibfnamefont {T.}~\bibnamefont
  {Rudolph}},\ }\bibfield  {title} {\bibinfo {title} {Proposal for pulsed
  on-demand sources of photonic cluster state strings},\ }\href
  {https://doi.org/10.1103/PhysRevLett.103.113602} {\bibfield  {journal}
  {\bibinfo  {journal} {Phys. Rev. Lett.}\ }\textbf {\bibinfo {volume} {103}},\
  \bibinfo {pages} {113602} (\bibinfo {year} {2009})}\BibitemShut {NoStop}%
\bibitem [{\citenamefont {Denning}\ \emph {et~al.}(2017)\citenamefont
  {Denning}, \citenamefont {Iles-Smith}, \citenamefont {McCutcheon},\ and\
  \citenamefont {Mork}}]{denning2017protocol}%
  \BibitemOpen
  \bibfield  {author} {\bibinfo {author} {\bibfnamefont {E.~V.}\ \bibnamefont
  {Denning}}, \bibinfo {author} {\bibfnamefont {J.}~\bibnamefont {Iles-Smith}},
  \bibinfo {author} {\bibfnamefont {D.~P.~S.}\ \bibnamefont {McCutcheon}},\
  and\ \bibinfo {author} {\bibfnamefont {J.}~\bibnamefont {Mork}},\ }\bibfield
  {title} {\bibinfo {title} {Protocol for generating multiphoton entangled
  states from quantum dots in the presence of nuclear spin fluctuations},\
  }\href {https://journals.aps.org/pra/abstract/10.1103/PhysRevA.96.062329}
  {\bibfield  {journal} {\bibinfo  {journal} {Phys. Rev. A}\ }\textbf {\bibinfo
  {volume} {96}},\ \bibinfo {pages} {062329} (\bibinfo {year}
  {2017})}\BibitemShut {NoStop}%
\bibitem [{\citenamefont {Scerri}\ \emph {et~al.}(2018)\citenamefont {Scerri},
  \citenamefont {Malein}, \citenamefont {Gerardot},\ and\ \citenamefont
  {Gauger}}]{scerri2018cluster}%
  \BibitemOpen
  \bibfield  {author} {\bibinfo {author} {\bibfnamefont {D.}~\bibnamefont
  {Scerri}}, \bibinfo {author} {\bibfnamefont {R.~N.~E.}\ \bibnamefont
  {Malein}}, \bibinfo {author} {\bibfnamefont {B.~D.}\ \bibnamefont
  {Gerardot}},\ and\ \bibinfo {author} {\bibfnamefont {E.~M.}\ \bibnamefont
  {Gauger}},\ }\bibfield  {title} {\bibinfo {title} {Frequency-encoded linear
  cluster states with coherent raman photons},\ }\href
  {https://doi.org/10.1103/PhysRevA.98.022318} {\bibfield  {journal} {\bibinfo
  {journal} {Phys. Rev. A}\ }\textbf {\bibinfo {volume} {98}},\ \bibinfo
  {pages} {022318} (\bibinfo {year} {2018})}\BibitemShut {NoStop}%
\bibitem [{\citenamefont {Schwartz}\ \emph {et~al.}(2016)\citenamefont
  {Schwartz}, \citenamefont {Cogan}, \citenamefont {Schmidgall}, \citenamefont
  {Don}, \citenamefont {Gantz}, \citenamefont {Kenneth}, \citenamefont
  {Lindner},\ and\ \citenamefont {Gershoni}}]{Schwartz434cluster}%
  \BibitemOpen
  \bibfield  {author} {\bibinfo {author} {\bibfnamefont {I.}~\bibnamefont
  {Schwartz}}, \bibinfo {author} {\bibfnamefont {D.}~\bibnamefont {Cogan}},
  \bibinfo {author} {\bibfnamefont {E.~R.}\ \bibnamefont {Schmidgall}},
  \bibinfo {author} {\bibfnamefont {Y.}~\bibnamefont {Don}}, \bibinfo {author}
  {\bibfnamefont {L.}~\bibnamefont {Gantz}}, \bibinfo {author} {\bibfnamefont
  {O.}~\bibnamefont {Kenneth}}, \bibinfo {author} {\bibfnamefont {N.~H.}\
  \bibnamefont {Lindner}},\ and\ \bibinfo {author} {\bibfnamefont
  {D.}~\bibnamefont {Gershoni}},\ }\bibfield  {title} {\bibinfo {title}
  {Deterministic generation of a cluster state of entangled photons},\ }\href
  {https://doi.org/10.1126/science.aah4758} {\bibfield  {journal} {\bibinfo
  {journal} {Science}\ }\textbf {\bibinfo {volume} {354}},\ \bibinfo {pages}
  {434} (\bibinfo {year} {2016})}\BibitemShut {NoStop}%
\bibitem [{\citenamefont {Cogan}\ \emph {et~al.}(2023)\citenamefont {Cogan},
  \citenamefont {Su}, \citenamefont {Kenneth},\ and\ \citenamefont
  {Gershoni}}]{cogan2023deterministic}%
  \BibitemOpen
  \bibfield  {author} {\bibinfo {author} {\bibfnamefont {D.}~\bibnamefont
  {Cogan}}, \bibinfo {author} {\bibfnamefont {Z.-E.}\ \bibnamefont {Su}},
  \bibinfo {author} {\bibfnamefont {O.}~\bibnamefont {Kenneth}},\ and\ \bibinfo
  {author} {\bibfnamefont {D.}~\bibnamefont {Gershoni}},\ }\bibfield  {title}
  {\bibinfo {title} {Deterministic generation of indistinguishable photons in a
  cluster state},\ }\href@noop {} {\bibfield  {journal} {\bibinfo  {journal}
  {Nature Photonics}\ }\textbf {\bibinfo {volume} {17}},\ \bibinfo {pages}
  {324} (\bibinfo {year} {2023})}\BibitemShut {NoStop}%
\bibitem [{\citenamefont {Coste}\ \emph {et~al.}(2023)\citenamefont {Coste}
  \emph {et~al.}}]{Coste:2022hfm}%
  \BibitemOpen
  \bibfield  {author} {\bibinfo {author} {\bibfnamefont {N.}~\bibnamefont
  {Coste}} \emph {et~al.},\ }\bibfield  {title} {\bibinfo {title} {{High-rate
  entanglement between a semiconductor spin and indistinguishable photons}},\
  }\href@noop {} {\bibfield  {journal} {\bibinfo  {journal} {Nature Photon.}\
  }\textbf {\bibinfo {volume} {17}},\ \bibinfo {pages} {582} (\bibinfo {year}
  {2023})}\BibitemShut {NoStop}%
\bibitem [{\citenamefont {Appel}\ \emph {et~al.}(2022)\citenamefont {Appel},
  \citenamefont {Tiranov}, \citenamefont {Pabst}, \citenamefont {Chan},
  \citenamefont {Starup}, \citenamefont {Wang}, \citenamefont {Midolo},
  \citenamefont {Tiurev}, \citenamefont {Scholz}, \citenamefont {Wieck},
  \citenamefont {Ludwig}, \citenamefont {S\o{}rensen},\ and\ \citenamefont
  {Lodahl}}]{PhysRevLett.128.233602}%
  \BibitemOpen
  \bibfield  {author} {\bibinfo {author} {\bibfnamefont {M.~H.}\ \bibnamefont
  {Appel}}, \bibinfo {author} {\bibfnamefont {A.}~\bibnamefont {Tiranov}},
  \bibinfo {author} {\bibfnamefont {S.}~\bibnamefont {Pabst}}, \bibinfo
  {author} {\bibfnamefont {M.~L.}\ \bibnamefont {Chan}}, \bibinfo {author}
  {\bibfnamefont {C.}~\bibnamefont {Starup}}, \bibinfo {author} {\bibfnamefont
  {Y.}~\bibnamefont {Wang}}, \bibinfo {author} {\bibfnamefont {L.}~\bibnamefont
  {Midolo}}, \bibinfo {author} {\bibfnamefont {K.}~\bibnamefont {Tiurev}},
  \bibinfo {author} {\bibfnamefont {S.}~\bibnamefont {Scholz}}, \bibinfo
  {author} {\bibfnamefont {A.~D.}\ \bibnamefont {Wieck}}, \bibinfo {author}
  {\bibfnamefont {A.}~\bibnamefont {Ludwig}}, \bibinfo {author} {\bibfnamefont
  {A.~S.}\ \bibnamefont {S\o{}rensen}},\ and\ \bibinfo {author} {\bibfnamefont
  {P.}~\bibnamefont {Lodahl}},\ }\bibfield  {title} {\bibinfo {title}
  {Entangling a hole spin with a time-bin photon: A waveguide approach for
  quantum dot sources of multiphoton entanglement},\ }\href
  {https://doi.org/10.1103/PhysRevLett.128.233602} {\bibfield  {journal}
  {\bibinfo  {journal} {Phys. Rev. Lett.}\ }\textbf {\bibinfo {volume} {128}},\
  \bibinfo {pages} {233602} (\bibinfo {year} {2022})}\BibitemShut {NoStop}%
\bibitem [{\citenamefont {Meng}\ \emph {et~al.}(2023)\citenamefont {Meng},
  \citenamefont {Faurby}, \citenamefont {Chan}, \citenamefont {Sund},
  \citenamefont {Liu}, \citenamefont {Wang}, \citenamefont {Bart},
  \citenamefont {Wieck}, \citenamefont {Ludwig}, \citenamefont {Midolo} \emph
  {et~al.}}]{meng2023photonic}%
  \BibitemOpen
  \bibfield  {author} {\bibinfo {author} {\bibfnamefont {Y.}~\bibnamefont
  {Meng}}, \bibinfo {author} {\bibfnamefont {C.~F.}\ \bibnamefont {Faurby}},
  \bibinfo {author} {\bibfnamefont {M.~L.}\ \bibnamefont {Chan}}, \bibinfo
  {author} {\bibfnamefont {P.~I.}\ \bibnamefont {Sund}}, \bibinfo {author}
  {\bibfnamefont {Z.}~\bibnamefont {Liu}}, \bibinfo {author} {\bibfnamefont
  {Y.}~\bibnamefont {Wang}}, \bibinfo {author} {\bibfnamefont {N.}~\bibnamefont
  {Bart}}, \bibinfo {author} {\bibfnamefont {A.~D.}\ \bibnamefont {Wieck}},
  \bibinfo {author} {\bibfnamefont {A.}~\bibnamefont {Ludwig}}, \bibinfo
  {author} {\bibfnamefont {L.}~\bibnamefont {Midolo}}, \emph {et~al.},\
  }\bibfield  {title} {\bibinfo {title} {Photonic fusion of entangled resource
  states from a quantum emitter},\ }\href@noop {} {\bibfield  {journal}
  {\bibinfo  {journal} {arXiv preprint arXiv:2312.09070}\ } (\bibinfo {year}
  {2023})}\BibitemShut {NoStop}%
\bibitem [{\citenamefont {Sun}\ \emph {et~al.}(2016)\citenamefont {Sun},
  \citenamefont {Kim}, \citenamefont {Solomon},\ and\ \citenamefont
  {Waks}}]{Sun2016}%
  \BibitemOpen
  \bibfield  {author} {\bibinfo {author} {\bibfnamefont {S.}~\bibnamefont
  {Sun}}, \bibinfo {author} {\bibfnamefont {H.}~\bibnamefont {Kim}}, \bibinfo
  {author} {\bibfnamefont {G.~S.}\ \bibnamefont {Solomon}},\ and\ \bibinfo
  {author} {\bibfnamefont {E.}~\bibnamefont {Waks}},\ }\bibfield  {title}
  {\bibinfo {title} {A quantum phase switch between a single solid-state spin
  and a photon},\ }\href {https://doi.org/10.1038/nnano.2015.334} {\bibfield
  {journal} {\bibinfo  {journal} {Nature Nanotechnology}\ }\textbf {\bibinfo
  {volume} {11}},\ \bibinfo {pages} {539} (\bibinfo {year} {2016})}\BibitemShut
  {NoStop}%
\bibitem [{\citenamefont {Sun}\ \emph {et~al.}(2018{\natexlab{a}})\citenamefont
  {Sun}, \citenamefont {Kim}, \citenamefont {Luo}, \citenamefont {Solomon},\
  and\ \citenamefont {Waks}}]{Sun57}%
  \BibitemOpen
  \bibfield  {author} {\bibinfo {author} {\bibfnamefont {S.}~\bibnamefont
  {Sun}}, \bibinfo {author} {\bibfnamefont {H.}~\bibnamefont {Kim}}, \bibinfo
  {author} {\bibfnamefont {Z.}~\bibnamefont {Luo}}, \bibinfo {author}
  {\bibfnamefont {G.~S.}\ \bibnamefont {Solomon}},\ and\ \bibinfo {author}
  {\bibfnamefont {E.}~\bibnamefont {Waks}},\ }\bibfield  {title} {\bibinfo
  {title} {A single-photon switch and transistor enabled by a solid-state
  quantum memory},\ }\href {https://doi.org/10.1126/science.aat3581} {\bibfield
   {journal} {\bibinfo  {journal} {Science}\ }\textbf {\bibinfo {volume}
  {361}},\ \bibinfo {pages} {57} (\bibinfo {year}
  {2018}{\natexlab{a}})}\BibitemShut {NoStop}%
\bibitem [{\citenamefont {Borregaard}\ \emph {et~al.}(2020)\citenamefont
  {Borregaard}, \citenamefont {Pichler}, \citenamefont {Schr\"oder},
  \citenamefont {Lukin}, \citenamefont {Lodahl},\ and\ \citenamefont
  {S\o{}rensen}}]{PhysRevX.10.021071}%
  \BibitemOpen
  \bibfield  {author} {\bibinfo {author} {\bibfnamefont {J.}~\bibnamefont
  {Borregaard}}, \bibinfo {author} {\bibfnamefont {H.}~\bibnamefont {Pichler}},
  \bibinfo {author} {\bibfnamefont {T.}~\bibnamefont {Schr\"oder}}, \bibinfo
  {author} {\bibfnamefont {M.~D.}\ \bibnamefont {Lukin}}, \bibinfo {author}
  {\bibfnamefont {P.}~\bibnamefont {Lodahl}},\ and\ \bibinfo {author}
  {\bibfnamefont {A.~S.}\ \bibnamefont {S\o{}rensen}},\ }\bibfield  {title}
  {\bibinfo {title} {One-way quantum repeater based on near-deterministic
  photon-emitter interfaces},\ }\href
  {https://doi.org/10.1103/PhysRevX.10.021071} {\bibfield  {journal} {\bibinfo
  {journal} {Phys. Rev. X}\ }\textbf {\bibinfo {volume} {10}},\ \bibinfo
  {pages} {021071} (\bibinfo {year} {2020})}\BibitemShut {NoStop}%
\bibitem [{\citenamefont {Munro}\ \emph {et~al.}(2012)\citenamefont {Munro},
  \citenamefont {Stephens}, \citenamefont {Devitt}, \citenamefont {Harrison},\
  and\ \citenamefont {Nemoto}}]{Munro2012}%
  \BibitemOpen
  \bibfield  {author} {\bibinfo {author} {\bibfnamefont {W.~J.}\ \bibnamefont
  {Munro}}, \bibinfo {author} {\bibfnamefont {A.~M.}\ \bibnamefont {Stephens}},
  \bibinfo {author} {\bibfnamefont {S.~J.}\ \bibnamefont {Devitt}}, \bibinfo
  {author} {\bibfnamefont {K.~A.}\ \bibnamefont {Harrison}},\ and\ \bibinfo
  {author} {\bibfnamefont {K.}~\bibnamefont {Nemoto}},\ }\bibfield  {title}
  {\bibinfo {title} {Quantum communication without the necessity of quantum
  memories},\ }\href {https://doi.org/10.1038/nphoton.2012.243} {\bibfield
  {journal} {\bibinfo  {journal} {Nature Photonics}\ }\textbf {\bibinfo
  {volume} {6}},\ \bibinfo {pages} {777} (\bibinfo {year} {2012})}\BibitemShut
  {NoStop}%
\bibitem [{\citenamefont {Atat{\"u}re}\ \emph {et~al.}(2006)\citenamefont
  {Atat{\"u}re}, \citenamefont {Dreiser}, \citenamefont {Badolato},
  \citenamefont {H{\"o}gele}, \citenamefont {Karrai},\ and\ \citenamefont
  {Imamoglu}}]{atature2006unity}%
  \BibitemOpen
  \bibfield  {author} {\bibinfo {author} {\bibfnamefont {M.}~\bibnamefont
  {Atat{\"u}re}}, \bibinfo {author} {\bibfnamefont {J.}~\bibnamefont
  {Dreiser}}, \bibinfo {author} {\bibfnamefont {A.}~\bibnamefont {Badolato}},
  \bibinfo {author} {\bibfnamefont {A.}~\bibnamefont {H{\"o}gele}}, \bibinfo
  {author} {\bibfnamefont {K.}~\bibnamefont {Karrai}},\ and\ \bibinfo {author}
  {\bibfnamefont {A.}~\bibnamefont {Imamoglu}},\ }\bibfield  {title} {\bibinfo
  {title} {Quantum-dot spin-state preparation with near-unity fidelity},\
  }\href {https://doi.org/10.1126/science.1126074} {\bibfield  {journal}
  {\bibinfo  {journal} {Science}\ }\textbf {\bibinfo {volume} {312}},\ \bibinfo
  {pages} {551} (\bibinfo {year} {2006})}\BibitemShut {NoStop}%
\bibitem [{\citenamefont {Kim}\ \emph {et~al.}(2008)\citenamefont {Kim},
  \citenamefont {Economou}, \citenamefont {B{\u{a}}descu}, \citenamefont
  {Scheibner}, \citenamefont {Bracker}, \citenamefont {Bashkansky},
  \citenamefont {Reinecke},\ and\ \citenamefont {Gammon}}]{Kim2008}%
  \BibitemOpen
  \bibfield  {author} {\bibinfo {author} {\bibfnamefont {D.}~\bibnamefont
  {Kim}}, \bibinfo {author} {\bibfnamefont {S.~E.}\ \bibnamefont {Economou}},
  \bibinfo {author} {\bibfnamefont {{\c{S}}.~C.}\ \bibnamefont
  {B{\u{a}}descu}}, \bibinfo {author} {\bibfnamefont {M.}~\bibnamefont
  {Scheibner}}, \bibinfo {author} {\bibfnamefont {A.~S.}\ \bibnamefont
  {Bracker}}, \bibinfo {author} {\bibfnamefont {M.}~\bibnamefont {Bashkansky}},
  \bibinfo {author} {\bibfnamefont {T.~L.}\ \bibnamefont {Reinecke}},\ and\
  \bibinfo {author} {\bibfnamefont {D.}~\bibnamefont {Gammon}},\ }\bibfield
  {title} {\bibinfo {title} {Optical spin initialization and nondestructive
  measurement in a quantum dot molecule},\ }\href
  {https://journals.aps.org/prl/abstract/10.1103/PhysRevLett.101.236804}
  {\bibfield  {journal} {\bibinfo  {journal} {Phys. Rev. Lett.}\ }\textbf
  {\bibinfo {volume} {101}},\ \bibinfo {pages} {236804} (\bibinfo {year}
  {2008})}\BibitemShut {NoStop}%
\bibitem [{Note1()}]{Note1}%
  \BibitemOpen
  \bibinfo {note} {A cycling transition is one which returns the system to its
  original spin state in a cyclical fashion.}\BibitemShut {Stop}%
\bibitem [{\citenamefont {Delteil}\ \emph {et~al.}(2014)\citenamefont
  {Delteil}, \citenamefont {Gao}, \citenamefont {Fallahi}, \citenamefont
  {Miguel-Sanchez},\ and\ \citenamefont {Imamo\ifmmode~\breve{g}\else
  \u{g}\fi{}lu}}]{PhysRevLett.112.116802}%
  \BibitemOpen
  \bibfield  {author} {\bibinfo {author} {\bibfnamefont {A.}~\bibnamefont
  {Delteil}}, \bibinfo {author} {\bibfnamefont {W.-b.}\ \bibnamefont {Gao}},
  \bibinfo {author} {\bibfnamefont {P.}~\bibnamefont {Fallahi}}, \bibinfo
  {author} {\bibfnamefont {J.}~\bibnamefont {Miguel-Sanchez}},\ and\ \bibinfo
  {author} {\bibfnamefont {A.}~\bibnamefont {Imamo\ifmmode~\breve{g}\else
  \u{g}\fi{}lu}},\ }\bibfield  {title} {\bibinfo {title} {Observation of
  quantum jumps of a single quantum dot spin using submicrosecond single-shot
  optical readout},\ }\href {https://doi.org/10.1103/PhysRevLett.112.116802}
  {\bibfield  {journal} {\bibinfo  {journal} {Phys. Rev. Lett.}\ }\textbf
  {\bibinfo {volume} {112}},\ \bibinfo {pages} {116802} (\bibinfo {year}
  {2014})}\BibitemShut {NoStop}%
\bibitem [{\citenamefont {Antoniadis}\ \emph {et~al.}(2023)\citenamefont
  {Antoniadis}, \citenamefont {Hogg}, \citenamefont {Stehl}, \citenamefont
  {Javadi}, \citenamefont {Tomm}, \citenamefont {Schott}, \citenamefont
  {Valentin}, \citenamefont {Wieck}, \citenamefont {Ludwig},\ and\
  \citenamefont {Warburton}}]{antoniadis2023cavity}%
  \BibitemOpen
  \bibfield  {author} {\bibinfo {author} {\bibfnamefont {N.~O.}\ \bibnamefont
  {Antoniadis}}, \bibinfo {author} {\bibfnamefont {M.~R.}\ \bibnamefont
  {Hogg}}, \bibinfo {author} {\bibfnamefont {W.~F.}\ \bibnamefont {Stehl}},
  \bibinfo {author} {\bibfnamefont {A.}~\bibnamefont {Javadi}}, \bibinfo
  {author} {\bibfnamefont {N.}~\bibnamefont {Tomm}}, \bibinfo {author}
  {\bibfnamefont {R.}~\bibnamefont {Schott}}, \bibinfo {author} {\bibfnamefont
  {S.~R.}\ \bibnamefont {Valentin}}, \bibinfo {author} {\bibfnamefont {A.~D.}\
  \bibnamefont {Wieck}}, \bibinfo {author} {\bibfnamefont {A.}~\bibnamefont
  {Ludwig}},\ and\ \bibinfo {author} {\bibfnamefont {R.~J.}\ \bibnamefont
  {Warburton}},\ }\bibfield  {title} {\bibinfo {title} {Cavity-enhanced
  single-shot readout of a quantum dot spin within 3 nanoseconds},\ }\href@noop
  {} {\bibfield  {journal} {\bibinfo  {journal} {Nature Communications}\
  }\textbf {\bibinfo {volume} {14}},\ \bibinfo {pages} {3977} (\bibinfo {year}
  {2023})}\BibitemShut {NoStop}%
\bibitem [{\citenamefont {Sun}\ and\ \citenamefont
  {Waks}(2016)}]{PhysRevA.94.012307}%
  \BibitemOpen
  \bibfield  {author} {\bibinfo {author} {\bibfnamefont {S.}~\bibnamefont
  {Sun}}\ and\ \bibinfo {author} {\bibfnamefont {E.}~\bibnamefont {Waks}},\
  }\bibfield  {title} {\bibinfo {title} {Single-shot optical readout of a
  quantum bit using cavity quantum electrodynamics},\ }\href
  {https://doi.org/10.1103/PhysRevA.94.012307} {\bibfield  {journal} {\bibinfo
  {journal} {Phys. Rev. A}\ }\textbf {\bibinfo {volume} {94}},\ \bibinfo
  {pages} {012307} (\bibinfo {year} {2016})}\BibitemShut {NoStop}%
\bibitem [{\citenamefont {Sun}\ \emph {et~al.}(2018{\natexlab{b}})\citenamefont
  {Sun}, \citenamefont {Kim}, \citenamefont {Solomon},\ and\ \citenamefont
  {Waks}}]{PhysRevApplied.9.054013}%
  \BibitemOpen
  \bibfield  {author} {\bibinfo {author} {\bibfnamefont {S.}~\bibnamefont
  {Sun}}, \bibinfo {author} {\bibfnamefont {H.}~\bibnamefont {Kim}}, \bibinfo
  {author} {\bibfnamefont {G.~S.}\ \bibnamefont {Solomon}},\ and\ \bibinfo
  {author} {\bibfnamefont {E.}~\bibnamefont {Waks}},\ }\bibfield  {title}
  {\bibinfo {title} {Cavity-enhanced optical readout of a single solid-state
  spin},\ }\href {https://doi.org/10.1103/PhysRevApplied.9.054013} {\bibfield
  {journal} {\bibinfo  {journal} {Phys. Rev. Applied}\ }\textbf {\bibinfo
  {volume} {9}},\ \bibinfo {pages} {054013} (\bibinfo {year}
  {2018}{\natexlab{b}})}\BibitemShut {NoStop}%
\bibitem [{\citenamefont {Appel}\ \emph {et~al.}(2021)\citenamefont {Appel},
  \citenamefont {Tiranov}, \citenamefont {Javadi}, \citenamefont {L\"obl},
  \citenamefont {Wang}, \citenamefont {Scholz}, \citenamefont {Wieck},
  \citenamefont {Ludwig}, \citenamefont {Warburton},\ and\ \citenamefont
  {Lodahl}}]{PhysRevLett.126.013602}%
  \BibitemOpen
  \bibfield  {author} {\bibinfo {author} {\bibfnamefont {M.~H.}\ \bibnamefont
  {Appel}}, \bibinfo {author} {\bibfnamefont {A.}~\bibnamefont {Tiranov}},
  \bibinfo {author} {\bibfnamefont {A.}~\bibnamefont {Javadi}}, \bibinfo
  {author} {\bibfnamefont {M.~C.}\ \bibnamefont {L\"obl}}, \bibinfo {author}
  {\bibfnamefont {Y.}~\bibnamefont {Wang}}, \bibinfo {author} {\bibfnamefont
  {S.}~\bibnamefont {Scholz}}, \bibinfo {author} {\bibfnamefont {A.~D.}\
  \bibnamefont {Wieck}}, \bibinfo {author} {\bibfnamefont {A.}~\bibnamefont
  {Ludwig}}, \bibinfo {author} {\bibfnamefont {R.~J.}\ \bibnamefont
  {Warburton}},\ and\ \bibinfo {author} {\bibfnamefont {P.}~\bibnamefont
  {Lodahl}},\ }\bibfield  {title} {\bibinfo {title} {Coherent spin-photon
  interface with waveguide induced cycling transitions},\ }\href
  {https://doi.org/10.1103/PhysRevLett.126.013602} {\bibfield  {journal}
  {\bibinfo  {journal} {Phys. Rev. Lett.}\ }\textbf {\bibinfo {volume} {126}},\
  \bibinfo {pages} {013602} (\bibinfo {year} {2021})}\BibitemShut {NoStop}%
\bibitem [{\citenamefont {Emary}\ \emph {et~al.}(2007)\citenamefont {Emary},
  \citenamefont {Xu}, \citenamefont {Steel}, \citenamefont {Saikin},\ and\
  \citenamefont {Sham}}]{emary2007fast}%
  \BibitemOpen
  \bibfield  {author} {\bibinfo {author} {\bibfnamefont {C.}~\bibnamefont
  {Emary}}, \bibinfo {author} {\bibfnamefont {X.}~\bibnamefont {Xu}}, \bibinfo
  {author} {\bibfnamefont {D.~G.}\ \bibnamefont {Steel}}, \bibinfo {author}
  {\bibfnamefont {S.}~\bibnamefont {Saikin}},\ and\ \bibinfo {author}
  {\bibfnamefont {L.~J.}\ \bibnamefont {Sham}},\ }\bibfield  {title} {\bibinfo
  {title} {Fast initialization of the spin state of an electron in a quantum
  dot in the voigt configuration},\ }\href
  {https://doi.org/10.1103/PhysRevLett.98.047401} {\bibfield  {journal}
  {\bibinfo  {journal} {Phys. Rev. Lett.}\ }\textbf {\bibinfo {volume} {98}},\
  \bibinfo {pages} {047401} (\bibinfo {year} {2007})}\BibitemShut {NoStop}%
\bibitem [{\citenamefont {Arcari}\ \emph {et~al.}(2014)\citenamefont {Arcari},
  \citenamefont {S\"ollner}, \citenamefont {Javadi}, \citenamefont
  {Lindskov~Hansen}, \citenamefont {Mahmoodian}, \citenamefont {Liu},
  \citenamefont {Thyrrestrup}, \citenamefont {Lee}, \citenamefont {Song},
  \citenamefont {Stobbe},\ and\ \citenamefont
  {Lodahl}}]{PhysRevLett.113.093603}%
  \BibitemOpen
  \bibfield  {author} {\bibinfo {author} {\bibfnamefont {M.}~\bibnamefont
  {Arcari}}, \bibinfo {author} {\bibfnamefont {I.}~\bibnamefont {S\"ollner}},
  \bibinfo {author} {\bibfnamefont {A.}~\bibnamefont {Javadi}}, \bibinfo
  {author} {\bibfnamefont {S.}~\bibnamefont {Lindskov~Hansen}}, \bibinfo
  {author} {\bibfnamefont {S.}~\bibnamefont {Mahmoodian}}, \bibinfo {author}
  {\bibfnamefont {J.}~\bibnamefont {Liu}}, \bibinfo {author} {\bibfnamefont
  {H.}~\bibnamefont {Thyrrestrup}}, \bibinfo {author} {\bibfnamefont {E.~H.}\
  \bibnamefont {Lee}}, \bibinfo {author} {\bibfnamefont {J.~D.}\ \bibnamefont
  {Song}}, \bibinfo {author} {\bibfnamefont {S.}~\bibnamefont {Stobbe}},\ and\
  \bibinfo {author} {\bibfnamefont {P.}~\bibnamefont {Lodahl}},\ }\bibfield
  {title} {\bibinfo {title} {Near-unity coupling efficiency of a quantum
  emitter to a photonic crystal waveguide},\ }\href
  {https://doi.org/10.1103/PhysRevLett.113.093603} {\bibfield  {journal}
  {\bibinfo  {journal} {Phys. Rev. Lett.}\ }\textbf {\bibinfo {volume} {113}},\
  \bibinfo {pages} {093603} (\bibinfo {year} {2014})}\BibitemShut {NoStop}%
\bibitem [{\citenamefont {Reinhard}\ \emph {et~al.}(2012)\citenamefont
  {Reinhard}, \citenamefont {Volz}, \citenamefont {Winger}, \citenamefont
  {Badolato}, \citenamefont {Hennessy}, \citenamefont {Hu},\ and\ \citenamefont
  {Imamoğlu}}]{Reinhard2012}%
  \BibitemOpen
  \bibfield  {author} {\bibinfo {author} {\bibfnamefont {A.}~\bibnamefont
  {Reinhard}}, \bibinfo {author} {\bibfnamefont {T.}~\bibnamefont {Volz}},
  \bibinfo {author} {\bibfnamefont {M.}~\bibnamefont {Winger}}, \bibinfo
  {author} {\bibfnamefont {A.}~\bibnamefont {Badolato}}, \bibinfo {author}
  {\bibfnamefont {K.~J.}\ \bibnamefont {Hennessy}}, \bibinfo {author}
  {\bibfnamefont {E.~L.}\ \bibnamefont {Hu}},\ and\ \bibinfo {author}
  {\bibfnamefont {A.}~\bibnamefont {Imamoğlu}},\ }\bibfield  {title} {\bibinfo
  {title} {Strongly correlated photons on a chip},\ }\href
  {https://doi.org/10.1038/nphoton.2011.321} {\bibfield  {journal} {\bibinfo
  {journal} {Nature Photonics}\ }\textbf {\bibinfo {volume} {6}},\ \bibinfo
  {pages} {93} (\bibinfo {year} {2012})}\BibitemShut {NoStop}%
\bibitem [{\citenamefont {Ohta}\ \emph {et~al.}(2011)\citenamefont {Ohta},
  \citenamefont {Ota}, \citenamefont {Nomura}, \citenamefont {Kumagai},
  \citenamefont {Ishida}, \citenamefont {Iwamoto},\ and\ \citenamefont
  {Arakawa}}]{doi:10.1063/1.3579535}%
  \BibitemOpen
  \bibfield  {author} {\bibinfo {author} {\bibfnamefont {R.}~\bibnamefont
  {Ohta}}, \bibinfo {author} {\bibfnamefont {Y.}~\bibnamefont {Ota}}, \bibinfo
  {author} {\bibfnamefont {M.}~\bibnamefont {Nomura}}, \bibinfo {author}
  {\bibfnamefont {N.}~\bibnamefont {Kumagai}}, \bibinfo {author} {\bibfnamefont
  {S.}~\bibnamefont {Ishida}}, \bibinfo {author} {\bibfnamefont
  {S.}~\bibnamefont {Iwamoto}},\ and\ \bibinfo {author} {\bibfnamefont
  {Y.}~\bibnamefont {Arakawa}},\ }\bibfield  {title} {\bibinfo {title} {Strong
  coupling between a photonic crystal nanobeam cavity and a single quantum
  dot},\ }\href {https://doi.org/10.1063/1.3579535} {\bibfield  {journal}
  {\bibinfo  {journal} {App. Phys. Lett.}\ }\textbf {\bibinfo {volume} {98}},\
  \bibinfo {pages} {173104} (\bibinfo {year} {2011})}\BibitemShut {NoStop}%
\bibitem [{\citenamefont {Reithmaier}\ \emph {et~al.}(2004)\citenamefont
  {Reithmaier}, \citenamefont {Sęk}, \citenamefont {Löffler}, \citenamefont
  {Hofmann}, \citenamefont {Kuhn}, \citenamefont {Reitzenstein}, \citenamefont
  {Keldysh}, \citenamefont {Kulakovskii}, \citenamefont {Reinecke},\ and\
  \citenamefont {Forchel}}]{Reithmaier2004}%
  \BibitemOpen
  \bibfield  {author} {\bibinfo {author} {\bibfnamefont {J.~P.}\ \bibnamefont
  {Reithmaier}}, \bibinfo {author} {\bibfnamefont {G.}~\bibnamefont {Sęk}},
  \bibinfo {author} {\bibfnamefont {A.}~\bibnamefont {Löffler}}, \bibinfo
  {author} {\bibfnamefont {C.}~\bibnamefont {Hofmann}}, \bibinfo {author}
  {\bibfnamefont {S.}~\bibnamefont {Kuhn}}, \bibinfo {author} {\bibfnamefont
  {S.}~\bibnamefont {Reitzenstein}}, \bibinfo {author} {\bibfnamefont {L.~V.}\
  \bibnamefont {Keldysh}}, \bibinfo {author} {\bibfnamefont {V.~D.}\
  \bibnamefont {Kulakovskii}}, \bibinfo {author} {\bibfnamefont {T.~L.}\
  \bibnamefont {Reinecke}},\ and\ \bibinfo {author} {\bibfnamefont
  {A.}~\bibnamefont {Forchel}},\ }\bibfield  {title} {\bibinfo {title} {Strong
  coupling in a single quantum dot-semiconductor microcavity system},\ }\href
  {https://doi.org/10.1038/nature02969} {\bibfield  {journal} {\bibinfo
  {journal} {Nature}\ }\textbf {\bibinfo {volume} {432}},\ \bibinfo {pages}
  {197} (\bibinfo {year} {2004})}\BibitemShut {NoStop}%
\bibitem [{\citenamefont {Liu}\ \emph {et~al.}(2018)\citenamefont {Liu},
  \citenamefont {Brash}, \citenamefont {O’Hara}, \citenamefont {Martins},
  \citenamefont {Phillips}, \citenamefont {Coles}, \citenamefont {Royall},
  \citenamefont {Clarke}, \citenamefont {Bentham}, \citenamefont {Prtljaga},
  \citenamefont {Itskevich}, \citenamefont {Wilson}, \citenamefont {Skolnick},\
  and\ \citenamefont {Fox}}]{Liu2018}%
  \BibitemOpen
  \bibfield  {author} {\bibinfo {author} {\bibfnamefont {F.}~\bibnamefont
  {Liu}}, \bibinfo {author} {\bibfnamefont {A.~J.}\ \bibnamefont {Brash}},
  \bibinfo {author} {\bibfnamefont {J.}~\bibnamefont {O’Hara}}, \bibinfo
  {author} {\bibfnamefont {L.~M. P.~P.}\ \bibnamefont {Martins}}, \bibinfo
  {author} {\bibfnamefont {C.~L.}\ \bibnamefont {Phillips}}, \bibinfo {author}
  {\bibfnamefont {R.~J.}\ \bibnamefont {Coles}}, \bibinfo {author}
  {\bibfnamefont {B.}~\bibnamefont {Royall}}, \bibinfo {author} {\bibfnamefont
  {E.}~\bibnamefont {Clarke}}, \bibinfo {author} {\bibfnamefont
  {C.}~\bibnamefont {Bentham}}, \bibinfo {author} {\bibfnamefont
  {N.}~\bibnamefont {Prtljaga}}, \bibinfo {author} {\bibfnamefont {I.~E.}\
  \bibnamefont {Itskevich}}, \bibinfo {author} {\bibfnamefont {L.~R.}\
  \bibnamefont {Wilson}}, \bibinfo {author} {\bibfnamefont {M.~S.}\
  \bibnamefont {Skolnick}},\ and\ \bibinfo {author} {\bibfnamefont {A.~M.}\
  \bibnamefont {Fox}},\ }\bibfield  {title} {\bibinfo {title} {High purcell
  factor generation of indistinguishable on-chip single photons},\ }\href
  {https://doi.org/10.1038/s41565-018-0188-x} {\bibfield  {journal} {\bibinfo
  {journal} {Nature Nanotechnology}\ }\textbf {\bibinfo {volume} {13}},\
  \bibinfo {pages} {835} (\bibinfo {year} {2018})}\BibitemShut {NoStop}%
\bibitem [{\citenamefont {Rivoire}\ \emph {et~al.}(2011)\citenamefont
  {Rivoire}, \citenamefont {Buckley},\ and\ \citenamefont
  {Vu{\v{c}}kovi{\'{c}}}}]{Rivoire2011}%
  \BibitemOpen
  \bibfield  {author} {\bibinfo {author} {\bibfnamefont {K.}~\bibnamefont
  {Rivoire}}, \bibinfo {author} {\bibfnamefont {S.}~\bibnamefont {Buckley}},\
  and\ \bibinfo {author} {\bibfnamefont {J.}~\bibnamefont
  {Vu{\v{c}}kovi{\'{c}}}},\ }\bibfield  {title} {\bibinfo {title} {{Multiply
  resonant high quality photonic crystal nanocavities}},\ }\href
  {https://doi.org/10.1063/1.3607281} {\bibfield  {journal} {\bibinfo
  {journal} {App. Phys. Lett.}\ }\textbf {\bibinfo {volume} {99}},\ \bibinfo
  {pages} {013114} (\bibinfo {year} {2011})}\BibitemShut {NoStop}%
\bibitem [{\citenamefont {Tomm}\ \emph {et~al.}(2021)\citenamefont {Tomm},
  \citenamefont {Javadi}, \citenamefont {Antoniadis}, \citenamefont {Najer},
  \citenamefont {L{\"{o}}bl}, \citenamefont {Korsch}, \citenamefont {Schott},
  \citenamefont {Valentin}, \citenamefont {Wieck}, \citenamefont {Ludwig},\
  and\ \citenamefont {Warburton}}]{Tomm2021bright}%
  \BibitemOpen
  \bibfield  {author} {\bibinfo {author} {\bibfnamefont {N.}~\bibnamefont
  {Tomm}}, \bibinfo {author} {\bibfnamefont {A.}~\bibnamefont {Javadi}},
  \bibinfo {author} {\bibfnamefont {N.~O.}\ \bibnamefont {Antoniadis}},
  \bibinfo {author} {\bibfnamefont {D.}~\bibnamefont {Najer}}, \bibinfo
  {author} {\bibfnamefont {M.~C.}\ \bibnamefont {L{\"{o}}bl}}, \bibinfo
  {author} {\bibfnamefont {A.~R.}\ \bibnamefont {Korsch}}, \bibinfo {author}
  {\bibfnamefont {R.}~\bibnamefont {Schott}}, \bibinfo {author} {\bibfnamefont
  {S.~R.}\ \bibnamefont {Valentin}}, \bibinfo {author} {\bibfnamefont {A.~D.}\
  \bibnamefont {Wieck}}, \bibinfo {author} {\bibfnamefont {A.}~\bibnamefont
  {Ludwig}},\ and\ \bibinfo {author} {\bibfnamefont {R.~J.}\ \bibnamefont
  {Warburton}},\ }\bibfield  {title} {\bibinfo {title} {{A bright and fast
  source of coherent single photons}},\ }\href
  {https://doi.org/10.1038/s41565-020-00831-x} {\bibfield  {journal} {\bibinfo
  {journal} {Nature Nanotechnology}\ }\textbf {\bibinfo {volume} {16}},\
  \bibinfo {pages} {399} (\bibinfo {year} {2021})},\ \Eprint
  {https://arxiv.org/abs/2007.12654} {2007.12654} \BibitemShut {NoStop}%
\bibitem [{\citenamefont {Wang}\ \emph {et~al.}(2019)\citenamefont {Wang},
  \citenamefont {He}, \citenamefont {Chung}, \citenamefont {Hu}, \citenamefont
  {Yu}, \citenamefont {Chen}, \citenamefont {Ding}, \citenamefont {Chen},
  \citenamefont {Qin}, \citenamefont {Yang} \emph {et~al.}}]{wang2019towards}%
  \BibitemOpen
  \bibfield  {author} {\bibinfo {author} {\bibfnamefont {H.}~\bibnamefont
  {Wang}}, \bibinfo {author} {\bibfnamefont {Y.-M.}\ \bibnamefont {He}},
  \bibinfo {author} {\bibfnamefont {T.-H.}\ \bibnamefont {Chung}}, \bibinfo
  {author} {\bibfnamefont {H.}~\bibnamefont {Hu}}, \bibinfo {author}
  {\bibfnamefont {Y.}~\bibnamefont {Yu}}, \bibinfo {author} {\bibfnamefont
  {S.}~\bibnamefont {Chen}}, \bibinfo {author} {\bibfnamefont {X.}~\bibnamefont
  {Ding}}, \bibinfo {author} {\bibfnamefont {M.-C.}\ \bibnamefont {Chen}},
  \bibinfo {author} {\bibfnamefont {J.}~\bibnamefont {Qin}}, \bibinfo {author}
  {\bibfnamefont {X.}~\bibnamefont {Yang}}, \emph {et~al.},\ }\bibfield
  {title} {\bibinfo {title} {Towards optimal single-photon sources from
  polarized microcavities},\ }\href@noop {} {\bibfield  {journal} {\bibinfo
  {journal} {Nature Photonics}\ }\textbf {\bibinfo {volume} {13}},\ \bibinfo
  {pages} {770} (\bibinfo {year} {2019})}\BibitemShut {NoStop}%
\bibitem [{\citenamefont {Coles}\ \emph {et~al.}(2014)\citenamefont {Coles},
  \citenamefont {Prtljaga}, \citenamefont {Royall}, \citenamefont {Luxmoore},
  \citenamefont {Fox},\ and\ \citenamefont {Skolnick}}]{Coles2014}%
  \BibitemOpen
  \bibfield  {author} {\bibinfo {author} {\bibfnamefont {R.~J.}\ \bibnamefont
  {Coles}}, \bibinfo {author} {\bibfnamefont {N.}~\bibnamefont {Prtljaga}},
  \bibinfo {author} {\bibfnamefont {B.}~\bibnamefont {Royall}}, \bibinfo
  {author} {\bibfnamefont {I.~J.}\ \bibnamefont {Luxmoore}}, \bibinfo {author}
  {\bibfnamefont {A.~M.}\ \bibnamefont {Fox}},\ and\ \bibinfo {author}
  {\bibfnamefont {M.~S.}\ \bibnamefont {Skolnick}},\ }\bibfield  {title}
  {\bibinfo {title} {{Waveguide-coupled photonic crystal cavity for quantum dot
  spin readout}},\ }\href {https://doi.org/10.1364/OE.22.002376} {\bibfield
  {journal} {\bibinfo  {journal} {Optics Express}\ }\textbf {\bibinfo {volume}
  {22}},\ \bibinfo {pages} {2376} (\bibinfo {year} {2014})}\BibitemShut
  {NoStop}%
\bibitem [{\citenamefont {G{\"{u}}r}\ \emph {et~al.}(2021)\citenamefont
  {G{\"{u}}r}, \citenamefont {Mattes}, \citenamefont {Arslanagi{\'{c}}},\ and\
  \citenamefont {Gregersen}}]{Gur2021}%
  \BibitemOpen
  \bibfield  {author} {\bibinfo {author} {\bibfnamefont {U.~M.}\ \bibnamefont
  {G{\"{u}}r}}, \bibinfo {author} {\bibfnamefont {M.}~\bibnamefont {Mattes}},
  \bibinfo {author} {\bibfnamefont {S.}~\bibnamefont {Arslanagi{\'{c}}}},\ and\
  \bibinfo {author} {\bibfnamefont {N.}~\bibnamefont {Gregersen}},\ }\bibfield
  {title} {\bibinfo {title} {{Elliptical micropillar cavity design for highly
  efficient polarized emission of single photons}},\ }\bibfield  {journal}
  {\bibinfo  {journal} {App. Phys. Lett.}\ }\textbf {\bibinfo {volume} {118}},\
  \href {https://doi.org/10.1063/5.0041565} {10.1063/5.0041565} (\bibinfo
  {year} {2021})\BibitemShut {NoStop}%
\bibitem [{\citenamefont {Gayral}\ \emph {et~al.}(1998)\citenamefont {Gayral},
  \citenamefont {G{\'{e}}rard}, \citenamefont {Legrand}, \citenamefont
  {Costard},\ and\ \citenamefont {Thierry-Mieg}}]{Gayral1998}%
  \BibitemOpen
  \bibfield  {author} {\bibinfo {author} {\bibfnamefont {B.}~\bibnamefont
  {Gayral}}, \bibinfo {author} {\bibfnamefont {J.~M.}\ \bibnamefont
  {G{\'{e}}rard}}, \bibinfo {author} {\bibfnamefont {B.}~\bibnamefont
  {Legrand}}, \bibinfo {author} {\bibfnamefont {E.}~\bibnamefont {Costard}},\
  and\ \bibinfo {author} {\bibfnamefont {V.}~\bibnamefont {Thierry-Mieg}},\
  }\bibfield  {title} {\bibinfo {title} {{Optical study of GaAs/AlAs pillar
  microcavities with elliptical cross section}},\ }\href
  {https://doi.org/10.1063/1.120582} {\bibfield  {journal} {\bibinfo  {journal}
  {App. Phys. Lett.}\ }\textbf {\bibinfo {volume} {72}},\ \bibinfo {pages}
  {1421} (\bibinfo {year} {1998})}\BibitemShut {NoStop}%
\bibitem [{\citenamefont {Reitzenstein}\ \emph {et~al.}(2007)\citenamefont
  {Reitzenstein}, \citenamefont {Hofmann}, \citenamefont {Gorbunov},
  \citenamefont {Strauß}, \citenamefont {Kwon}, \citenamefont {Schneider},
  \citenamefont {Löffler}, \citenamefont {Höfling}, \citenamefont {Kamp},\
  and\ \citenamefont {Forchel}}]{doi:10.1063/1.2749862}%
  \BibitemOpen
  \bibfield  {author} {\bibinfo {author} {\bibfnamefont {S.}~\bibnamefont
  {Reitzenstein}}, \bibinfo {author} {\bibfnamefont {C.}~\bibnamefont
  {Hofmann}}, \bibinfo {author} {\bibfnamefont {A.}~\bibnamefont {Gorbunov}},
  \bibinfo {author} {\bibfnamefont {M.}~\bibnamefont {Strauß}}, \bibinfo
  {author} {\bibfnamefont {S.~H.}\ \bibnamefont {Kwon}}, \bibinfo {author}
  {\bibfnamefont {C.}~\bibnamefont {Schneider}}, \bibinfo {author}
  {\bibfnamefont {A.}~\bibnamefont {Löffler}}, \bibinfo {author}
  {\bibfnamefont {S.}~\bibnamefont {Höfling}}, \bibinfo {author}
  {\bibfnamefont {M.}~\bibnamefont {Kamp}},\ and\ \bibinfo {author}
  {\bibfnamefont {A.}~\bibnamefont {Forchel}},\ }\bibfield  {title} {\bibinfo
  {title} {Alas∕gaas micropillar cavities with quality factors exceeding
  150.000},\ }\href {https://doi.org/10.1063/1.2749862} {\bibfield  {journal}
  {\bibinfo  {journal} {App. Phys. Lett.}\ }\textbf {\bibinfo {volume} {90}},\
  \bibinfo {pages} {251109} (\bibinfo {year} {2007})}\BibitemShut {NoStop}%
\bibitem [{\citenamefont {Luxmoore}\ \emph {et~al.}(2012)\citenamefont
  {Luxmoore}, \citenamefont {Ahmadi}, \citenamefont {Luxmoore}, \citenamefont
  {Wasley}, \citenamefont {Tartakovskii}, \citenamefont {Hugues}, \citenamefont
  {Skolnick},\ and\ \citenamefont {Fox}}]{doi:10.1063/1.3696036}%
  \BibitemOpen
  \bibfield  {author} {\bibinfo {author} {\bibfnamefont {I.~J.}\ \bibnamefont
  {Luxmoore}}, \bibinfo {author} {\bibfnamefont {E.~D.}\ \bibnamefont
  {Ahmadi}}, \bibinfo {author} {\bibfnamefont {B.~J.}\ \bibnamefont
  {Luxmoore}}, \bibinfo {author} {\bibfnamefont {N.~A.}\ \bibnamefont
  {Wasley}}, \bibinfo {author} {\bibfnamefont {A.~I.}\ \bibnamefont
  {Tartakovskii}}, \bibinfo {author} {\bibfnamefont {M.}~\bibnamefont
  {Hugues}}, \bibinfo {author} {\bibfnamefont {M.~S.}\ \bibnamefont
  {Skolnick}},\ and\ \bibinfo {author} {\bibfnamefont {A.~M.}\ \bibnamefont
  {Fox}},\ }\bibfield  {title} {\bibinfo {title} {Restoring mode degeneracy in
  h1 photonic crystal cavities by uniaxial strain tuning},\ }\href
  {https://doi.org/10.1063/1.3696036} {\bibfield  {journal} {\bibinfo
  {journal} {App. Phys. Lett.}\ }\textbf {\bibinfo {volume} {100}},\ \bibinfo
  {pages} {121116} (\bibinfo {year} {2012})}\BibitemShut {NoStop}%
\bibitem [{\citenamefont {Carmichael}(1999)}]{carmichael1999statistical}%
  \BibitemOpen
  \bibfield  {author} {\bibinfo {author} {\bibfnamefont {H.~J.}\ \bibnamefont
  {Carmichael}},\ }\href@noop {} {\emph {\bibinfo {title} {Statistical methods
  in quantum optics 1: master equations and Fokker-Planck equations}}},\
  Vol.~\bibinfo {volume} {1}\ (\bibinfo  {publisher} {Springer Science \&
  Business Media},\ \bibinfo {year} {1999})\BibitemShut {NoStop}%
\bibitem [{\citenamefont {Johansson}\ \emph {et~al.}(2012)\citenamefont
  {Johansson}, \citenamefont {Nation},\ and\ \citenamefont
  {Nori}}]{johansson2012qutip}%
  \BibitemOpen
  \bibfield  {author} {\bibinfo {author} {\bibfnamefont {J.~R.}\ \bibnamefont
  {Johansson}}, \bibinfo {author} {\bibfnamefont {P.~D.}\ \bibnamefont
  {Nation}},\ and\ \bibinfo {author} {\bibfnamefont {F.}~\bibnamefont {Nori}},\
  }\bibfield  {title} {\bibinfo {title} {Qutip: An open-source python framework
  for the dynamics of open quantum systems},\ }\href
  {https://www.sciencedirect.com/science/article/pii/S0010465512000835}
  {\bibfield  {journal} {\bibinfo  {journal} {Computer Physics Communications}\
  }\textbf {\bibinfo {volume} {183}},\ \bibinfo {pages} {1760} (\bibinfo {year}
  {2012})}\BibitemShut {NoStop}%
\bibitem [{\citenamefont {Majumdar}\ \emph {et~al.}(2013)\citenamefont
  {Majumdar}, \citenamefont {Kaer}, \citenamefont {Bajcsy}, \citenamefont
  {Kim}, \citenamefont {Lagoudakis}, \citenamefont {Rundquist},\ and\
  \citenamefont {Vu\ifmmode \check{c}\else
  \v{c}\fi{}kovi\ifmmode~\acute{c}\else \'{c}\fi{}}}]{PhysRevLett.111.027402}%
  \BibitemOpen
  \bibfield  {author} {\bibinfo {author} {\bibfnamefont {A.}~\bibnamefont
  {Majumdar}}, \bibinfo {author} {\bibfnamefont {P.}~\bibnamefont {Kaer}},
  \bibinfo {author} {\bibfnamefont {M.}~\bibnamefont {Bajcsy}}, \bibinfo
  {author} {\bibfnamefont {E.~D.}\ \bibnamefont {Kim}}, \bibinfo {author}
  {\bibfnamefont {K.~G.}\ \bibnamefont {Lagoudakis}}, \bibinfo {author}
  {\bibfnamefont {A.}~\bibnamefont {Rundquist}},\ and\ \bibinfo {author}
  {\bibfnamefont {J.}~\bibnamefont {Vu\ifmmode \check{c}\else
  \v{c}\fi{}kovi\ifmmode~\acute{c}\else \'{c}\fi{}}},\ }\bibfield  {title}
  {\bibinfo {title} {Proposed coupling of an electron spin in a semiconductor
  quantum dot to a nanosize optical cavity},\ }\href
  {https://doi.org/10.1103/PhysRevLett.111.027402} {\bibfield  {journal}
  {\bibinfo  {journal} {Phys. Rev. Lett.}\ }\textbf {\bibinfo {volume} {111}},\
  \bibinfo {pages} {027402} (\bibinfo {year} {2013})}\BibitemShut {NoStop}%
\bibitem [{\citenamefont {Loo}\ \emph {et~al.}(2011)\citenamefont {Loo},
  \citenamefont {Lanco}, \citenamefont {Krebs}, \citenamefont {Senellart},\
  and\ \citenamefont {Voisin}}]{Loo2011}%
  \BibitemOpen
  \bibfield  {author} {\bibinfo {author} {\bibfnamefont {V.}~\bibnamefont
  {Loo}}, \bibinfo {author} {\bibfnamefont {L.}~\bibnamefont {Lanco}}, \bibinfo
  {author} {\bibfnamefont {O.}~\bibnamefont {Krebs}}, \bibinfo {author}
  {\bibfnamefont {P.}~\bibnamefont {Senellart}},\ and\ \bibinfo {author}
  {\bibfnamefont {P.}~\bibnamefont {Voisin}},\ }\bibfield  {title} {\bibinfo
  {title} {{Single-shot initialization of electron spin in a quantum dot using
  a short optical pulse}},\ }\href {https://doi.org/10.1103/PhysRevB.83.033301}
  {\bibfield  {journal} {\bibinfo  {journal} {Phys. Rev. B}\ }\textbf {\bibinfo
  {volume} {83}},\ \bibinfo {pages} {033301} (\bibinfo {year} {2011})},\
  \Eprint {https://arxiv.org/abs/1011.1156} {1011.1156} \BibitemShut {NoStop}%
\bibitem [{\citenamefont {Gilchrist}\ \emph {et~al.}(2005)\citenamefont
  {Gilchrist}, \citenamefont {Langford},\ and\ \citenamefont
  {Nielsen}}]{Gilchrist2005}%
  \BibitemOpen
  \bibfield  {author} {\bibinfo {author} {\bibfnamefont {A.}~\bibnamefont
  {Gilchrist}}, \bibinfo {author} {\bibfnamefont {N.~K.}\ \bibnamefont
  {Langford}},\ and\ \bibinfo {author} {\bibfnamefont {M.~A.}\ \bibnamefont
  {Nielsen}},\ }\bibfield  {title} {\bibinfo {title} {Distance measures to
  compare real and ideal quantum processes},\ }\href@noop {} {\bibfield
  {journal} {\bibinfo  {journal} {Phys. Rev. A}\ }\textbf {\bibinfo {volume}
  {71}},\ \bibinfo {pages} {062310} (\bibinfo {year} {2005})}\BibitemShut
  {NoStop}%
\bibitem [{\citenamefont {Paspalakis}\ \emph {et~al.}(2019)\citenamefont
  {Paspalakis}, \citenamefont {Economou},\ and\ \citenamefont
  {Carre{\~{n}}o}}]{Paspalakis2019}%
  \BibitemOpen
  \bibfield  {author} {\bibinfo {author} {\bibfnamefont {E.}~\bibnamefont
  {Paspalakis}}, \bibinfo {author} {\bibfnamefont {S.~E.}\ \bibnamefont
  {Economou}},\ and\ \bibinfo {author} {\bibfnamefont {F.}~\bibnamefont
  {Carre{\~{n}}o}},\ }\bibfield  {title} {\bibinfo {title} {{Adiabatically
  preparing quantum dot spin states in the Voigt geometry}},\ }\href
  {http://dx.doi.org/10.1063/1.5079412} {\bibfield  {journal} {\bibinfo
  {journal} {Journal of Applied Physics}\ }\textbf {\bibinfo {volume} {125}}
  (\bibinfo {year} {2019})}\BibitemShut {NoStop}%
\bibitem [{\citenamefont {Gerardot}\ \emph {et~al.}(2008)\citenamefont
  {Gerardot}, \citenamefont {Brunner}, \citenamefont {Dalgarno}, \citenamefont
  {{\"{O}}hberg}, \citenamefont {Seidl}, \citenamefont {Kroner}, \citenamefont
  {Karrai}, \citenamefont {Stoltz}, \citenamefont {Petroff},\ and\
  \citenamefont {Warburton}}]{Gerardot2008}%
  \BibitemOpen
  \bibfield  {author} {\bibinfo {author} {\bibfnamefont {B.~D.}\ \bibnamefont
  {Gerardot}}, \bibinfo {author} {\bibfnamefont {D.}~\bibnamefont {Brunner}},
  \bibinfo {author} {\bibfnamefont {P.~A.}\ \bibnamefont {Dalgarno}}, \bibinfo
  {author} {\bibfnamefont {P.}~\bibnamefont {{\"{O}}hberg}}, \bibinfo {author}
  {\bibfnamefont {S.}~\bibnamefont {Seidl}}, \bibinfo {author} {\bibfnamefont
  {M.}~\bibnamefont {Kroner}}, \bibinfo {author} {\bibfnamefont
  {K.}~\bibnamefont {Karrai}}, \bibinfo {author} {\bibfnamefont {N.~G.}\
  \bibnamefont {Stoltz}}, \bibinfo {author} {\bibfnamefont {P.~M.}\
  \bibnamefont {Petroff}},\ and\ \bibinfo {author} {\bibfnamefont {R.~J.}\
  \bibnamefont {Warburton}},\ }\bibfield  {title} {\bibinfo {title} {{Optical
  pumping of a single hole spin in a quantum dot}},\ }\href
  {https://doi.org/10.1038/nature06472} {\bibfield  {journal} {\bibinfo
  {journal} {Nature}\ }\textbf {\bibinfo {volume} {451}},\ \bibinfo {pages}
  {441} (\bibinfo {year} {2008})}\BibitemShut {NoStop}%
\bibitem [{\citenamefont {Stefanatos}\ \emph {et~al.}(2019)\citenamefont
  {Stefanatos}, \citenamefont {Iliopoulos}, \citenamefont {Karanikolas},\ and\
  \citenamefont {Paspalakis}}]{Stefanatos2019}%
  \BibitemOpen
  \bibfield  {author} {\bibinfo {author} {\bibfnamefont {D.}~\bibnamefont
  {Stefanatos}}, \bibinfo {author} {\bibfnamefont {N.}~\bibnamefont
  {Iliopoulos}}, \bibinfo {author} {\bibfnamefont {V.}~\bibnamefont
  {Karanikolas}},\ and\ \bibinfo {author} {\bibfnamefont {E.}~\bibnamefont
  {Paspalakis}},\ }\bibfield  {title} {\bibinfo {title} {{Adiabatic control of
  quantum dot spin in the voigt geometry with optical pulses}},\ }\href
  {https://www.wseas.org/multimedia/journals/control/2019/a825103-076.php}
  {\bibfield  {journal} {\bibinfo  {journal} {WSEAS Transactions on Systems and
  Control}\ }\textbf {\bibinfo {volume} {14}},\ \bibinfo {pages} {319}
  (\bibinfo {year} {2019})}\BibitemShut {NoStop}%
\bibitem [{\citenamefont {Kumar}\ and\ \citenamefont
  {Nakajima}(2016)}]{Kumar2016}%
  \BibitemOpen
  \bibfield  {author} {\bibinfo {author} {\bibfnamefont {P.}~\bibnamefont
  {Kumar}}\ and\ \bibinfo {author} {\bibfnamefont {T.}~\bibnamefont
  {Nakajima}},\ }\bibfield  {title} {\bibinfo {title} {{Fast and high-fidelity
  optical initialization of spin state of an electron in a semiconductor
  quantum dot using light-hole-trion states}},\ }\href
  {https://doi.org/10.1016/j.optcom.2016.03.006} {\bibfield  {journal}
  {\bibinfo  {journal} {Optics Communications}\ }\textbf {\bibinfo {volume}
  {370}},\ \bibinfo {pages} {103} (\bibinfo {year} {2016})}\BibitemShut
  {NoStop}%
\bibitem [{Note2()}]{Note2}%
  \BibitemOpen
  \bibinfo {note} {In the CW driving regime, the two driving regimes are
  unitarily equivalent.}\BibitemShut {Stop}%
\bibitem [{Note3()}]{Note3}%
  \BibitemOpen
  \bibinfo {note} {We use a 35 ns readout pulse as a compromise to maximise the
  readout fidelity across a range of cavity parameters.}\BibitemShut {Stop}%
\bibitem [{\citenamefont {Tiecke}\ \emph {et~al.}(2015)\citenamefont {Tiecke},
  \citenamefont {Nayak}, \citenamefont {Thompson}, \citenamefont {Peyronel},
  \citenamefont {de~Leon}, \citenamefont {Vuleti{\'{c}}},\ and\ \citenamefont
  {Lukin}}]{Tiecke2015}%
  \BibitemOpen
  \bibfield  {author} {\bibinfo {author} {\bibfnamefont {T.~G.}\ \bibnamefont
  {Tiecke}}, \bibinfo {author} {\bibfnamefont {K.~P.}\ \bibnamefont {Nayak}},
  \bibinfo {author} {\bibfnamefont {J.~D.}\ \bibnamefont {Thompson}}, \bibinfo
  {author} {\bibfnamefont {T.}~\bibnamefont {Peyronel}}, \bibinfo {author}
  {\bibfnamefont {N.~P.}\ \bibnamefont {de~Leon}}, \bibinfo {author}
  {\bibfnamefont {V.}~\bibnamefont {Vuleti{\'{c}}}},\ and\ \bibinfo {author}
  {\bibfnamefont {M.~D.}\ \bibnamefont {Lukin}},\ }\bibfield  {title} {\bibinfo
  {title} {{Efficient fiber-optical interface for nanophotonic devices}},\
  }\href {https://doi.org/10.1364/OPTICA.2.000070} {\bibfield  {journal}
  {\bibinfo  {journal} {Optica}\ }\textbf {\bibinfo {volume} {2}},\ \bibinfo
  {pages} {70} (\bibinfo {year} {2015})},\ \Eprint
  {https://arxiv.org/abs/1409.7698} {1409.7698} \BibitemShut {NoStop}%
\bibitem [{\citenamefont {Daveau}\ \emph {et~al.}(2017)\citenamefont {Daveau},
  \citenamefont {Balram}, \citenamefont {Pregnolato}, \citenamefont {Liu},
  \citenamefont {Lee}, \citenamefont {Song}, \citenamefont {Verma},
  \citenamefont {Mirin}, \citenamefont {Nam}, \citenamefont {Midolo},
  \citenamefont {Stobbe}, \citenamefont {Srinivasan},\ and\ \citenamefont
  {Lodahl}}]{Daveau:17}%
  \BibitemOpen
  \bibfield  {author} {\bibinfo {author} {\bibfnamefont {R.~S.}\ \bibnamefont
  {Daveau}}, \bibinfo {author} {\bibfnamefont {K.~C.}\ \bibnamefont {Balram}},
  \bibinfo {author} {\bibfnamefont {T.}~\bibnamefont {Pregnolato}}, \bibinfo
  {author} {\bibfnamefont {J.}~\bibnamefont {Liu}}, \bibinfo {author}
  {\bibfnamefont {E.~H.}\ \bibnamefont {Lee}}, \bibinfo {author} {\bibfnamefont
  {J.~D.}\ \bibnamefont {Song}}, \bibinfo {author} {\bibfnamefont
  {V.}~\bibnamefont {Verma}}, \bibinfo {author} {\bibfnamefont
  {R.}~\bibnamefont {Mirin}}, \bibinfo {author} {\bibfnamefont {S.~W.}\
  \bibnamefont {Nam}}, \bibinfo {author} {\bibfnamefont {L.}~\bibnamefont
  {Midolo}}, \bibinfo {author} {\bibfnamefont {S.}~\bibnamefont {Stobbe}},
  \bibinfo {author} {\bibfnamefont {K.}~\bibnamefont {Srinivasan}},\ and\
  \bibinfo {author} {\bibfnamefont {P.}~\bibnamefont {Lodahl}},\ }\bibfield
  {title} {\bibinfo {title} {Efficient fiber-coupled single-photon source based
  on quantum dots in a photonic-crystal waveguide},\ }\href
  {https://doi.org/10.1364/OPTICA.4.000178} {\bibfield  {journal} {\bibinfo
  {journal} {Optica}\ }\textbf {\bibinfo {volume} {4}},\ \bibinfo {pages} {178}
  (\bibinfo {year} {2017})}\BibitemShut {NoStop}%
\bibitem [{\citenamefont {Volz}\ \emph {et~al.}(2012)\citenamefont {Volz},
  \citenamefont {Reinhard}, \citenamefont {Winger}, \citenamefont {Badolato},
  \citenamefont {Hennessy}, \citenamefont {Hu},\ and\ \citenamefont
  {Imamoğlu}}]{Volz2012}%
  \BibitemOpen
  \bibfield  {author} {\bibinfo {author} {\bibfnamefont {T.}~\bibnamefont
  {Volz}}, \bibinfo {author} {\bibfnamefont {A.}~\bibnamefont {Reinhard}},
  \bibinfo {author} {\bibfnamefont {M.}~\bibnamefont {Winger}}, \bibinfo
  {author} {\bibfnamefont {A.}~\bibnamefont {Badolato}}, \bibinfo {author}
  {\bibfnamefont {K.~J.}\ \bibnamefont {Hennessy}}, \bibinfo {author}
  {\bibfnamefont {E.~L.}\ \bibnamefont {Hu}},\ and\ \bibinfo {author}
  {\bibfnamefont {A.}~\bibnamefont {Imamoğlu}},\ }\bibfield  {title} {\bibinfo
  {title} {Ultrafast all-optical switching by single photons},\ }\href
  {https://doi.org/10.1038/nphoton.2012.181} {\bibfield  {journal} {\bibinfo
  {journal} {Nature Photonics}\ }\textbf {\bibinfo {volume} {6}},\ \bibinfo
  {pages} {605} (\bibinfo {year} {2012})}\BibitemShut {NoStop}%
\bibitem [{\citenamefont {Kuruma}\ \emph {et~al.}(2018)\citenamefont {Kuruma},
  \citenamefont {Ota}, \citenamefont {Kakuda}, \citenamefont {Iwamoto},\ and\
  \citenamefont {Arakawa}}]{PhysRevB.97.235448}%
  \BibitemOpen
  \bibfield  {author} {\bibinfo {author} {\bibfnamefont {K.}~\bibnamefont
  {Kuruma}}, \bibinfo {author} {\bibfnamefont {Y.}~\bibnamefont {Ota}},
  \bibinfo {author} {\bibfnamefont {M.}~\bibnamefont {Kakuda}}, \bibinfo
  {author} {\bibfnamefont {S.}~\bibnamefont {Iwamoto}},\ and\ \bibinfo {author}
  {\bibfnamefont {Y.}~\bibnamefont {Arakawa}},\ }\bibfield  {title} {\bibinfo
  {title} {Time-resolved vacuum rabi oscillations in a quantum-dot--nanocavity
  system},\ }\href {https://doi.org/10.1103/PhysRevB.97.235448} {\bibfield
  {journal} {\bibinfo  {journal} {Phys. Rev. B}\ }\textbf {\bibinfo {volume}
  {97}},\ \bibinfo {pages} {235448} (\bibinfo {year} {2018})}\BibitemShut
  {NoStop}%
\bibitem [{\citenamefont {Kuruma}\ \emph {et~al.}(2020)\citenamefont {Kuruma},
  \citenamefont {Ota}, \citenamefont {Kakuda}, \citenamefont {Iwamoto},\ and\
  \citenamefont {Arakawa}}]{doi:10.1063/1.5144959}%
  \BibitemOpen
  \bibfield  {author} {\bibinfo {author} {\bibfnamefont {K.}~\bibnamefont
  {Kuruma}}, \bibinfo {author} {\bibfnamefont {Y.}~\bibnamefont {Ota}},
  \bibinfo {author} {\bibfnamefont {M.}~\bibnamefont {Kakuda}}, \bibinfo
  {author} {\bibfnamefont {S.}~\bibnamefont {Iwamoto}},\ and\ \bibinfo {author}
  {\bibfnamefont {Y.}~\bibnamefont {Arakawa}},\ }\bibfield  {title} {\bibinfo
  {title} {Surface-passivated high-q gaas photonic crystal nanocavity with
  quantum dots},\ }\href {https://doi.org/10.1063/1.5144959} {\bibfield
  {journal} {\bibinfo  {journal} {APL Photonics}\ }\textbf {\bibinfo {volume}
  {5}},\ \bibinfo {pages} {046106} (\bibinfo {year} {2020})}\BibitemShut
  {NoStop}%
\end{thebibliography}%
\appendix
\end{document}